\documentclass[a4paper,fleqn]{cas-sc}

\usepackage{bbm}
\usepackage{amssymb,amsmath}
\usepackage{siunitx}
\usepackage[acronym,shortcuts]{glossaries}
\usepackage[font=footnotesize,labelfont=bf]{subcaption}

\usepackage[numbers]{natbib}

\newacronym{6g}{6G}{Sixth Generation}
\newacronym{cpu}{CPU}{Central Processing Unit}
\newacronym{fpga}{FPGA}{Field Programmable Gate Arrays}
\newacronym{sla}{SLA}{Service Level Agreement}
\newacronym{Sim-Diasca}{Sim-Diasca}{Simulation of Discrete Systems of All Scales}
\newacronym{capex}{CapEx}{Capital Expenditures}
\newacronym{opex}{OpEx}{Operational Expenditures}
\newacronym{zsm}{ZSM}{Zero-touch network and Service Management}
\newacronym{mape-k}{MAPE-K}{Monitor-Analyze-Plan-Execute over a shared Knowledge}
\newacronym{k8s}{K8s}{Kubernetes}
\newacronym{mlops}{MLOps}{Machine Learning Operations}
\newacronym{ndt}{NDT}{Network Digital Twin}
\newacronym{nif}{NIF}{Network Intelligent Function}
\newacronym{nis}{NIS}{Network Intelligent Service}
\newacronym{ni}{NI}{Network Intelligence}
\newacronym{ai}{AI}{Artificial Intelligence}
\newacronym{ml}{ML}{Machine Learning}
\newacronym{dl}{DL}{Deep Learning}
\newacronym{rl}{RL}{Reinforcement Learning}
\newacronym{drl}{DRL}{Deep Reinforcement Learning}
\newacronym{rfn}{RFn}{Reward Function}
\newacronym{dqn}{DQN}{Deep Q-Network}
\newacronym{ppo}{PPO}{Proximal Policy Optimization}
\newacronym{trpo}{TRPO}{Trust Region Policy Optimization}
\newacronym[plural=MDPs,firstplural=Markov Decision Processes (MDPs)]{mdp}{MDP}{Markov Decision Process}
\newacronym[plural=MRPs,firstplural=Markov Reward Processes (MRPs)]{mrp}{MRP}{Markov Reward Process}
\newacronym{rlops}{RLOps}{Reinforcement Learning Operations}
\newacronym{sl}{SL}{Supervised Learning}
\newacronym{gnn}{GNN}{Graph Neural Network}
\newacronym{ran}{RAN}{Radio Access Network}
\newacronym{sdn}{SDN}{Software Defined Networking}
\newacronym{nfv}{NFV}{Network Function Virtualization}
\newacronym{cots}{COTS}{Commercial Off-The-Shelf}
\newacronym{mano}{MANO}{Management and Orchestration}
\newacronym{nf}{NF}{Network Function}
\newacronym{vnf}{VNF}{Virtual Network Function}
\newacronym{ns}{NS}{Network Service}
\newacronym{nsp}{NSP}{Network Service Provider}
\newacronym{hpa}{HPA}{Horizontal Pod Autoscaler}
\newacronym{qos}{QoS}{Quality of Service}
\newacronym{qoe}{QoE}{Quality of Experience}
\newacronym{kpi}{KPI}{Key Performance Indicator}
\newacronym{mape}{MAPE}{Mean Average Percentage Error}
\newacronym{oran}{O-RAN}{Open Radio Access Network}
\newacronym{sb3}{SB3}{Stable-Baselines3}
\newacronym{0mq}{0MQ}{ZeroMQ}
\newacronym{protobuf}{Protobuf}{Google Protocol Buffers}
\newacronym{nio}{NIO}{Network Intelligence Orchestrator}

\def\tsc#1{\csdef{#1}{\textsc{\lowercase{#1}}\xspace}}
\tsc{WGM}
\tsc{QE}

\begin{document}
\let\WriteBookmarks\relax
\def\floatpagepagefraction{1}
\def\textpagefraction{.001}

\shorttitle{Designing, Developing, and Validating NI for scaling using DRL}    

\shortauthors{P. Soto et~al.}  

\title [mode = title]{Designing, Developing, and Validating Network Intelligence for Scaling in Service-Based Architectures based on Deep Reinforcement Learning}  



%

\author[a,b]{Paola~Soto}[orcid=0000-0000-0000-0000]
\cormark[1]
\ead{paola.soto-arenas@uantwerpen.be} 
\credit{Writing - Original Draft, Writing - Review \& Editing, Conceptualization, Methodology, Investigation, Software}
\author[a]{Miguel~Camelo}
\credit{Writing - Review \& Editing, Methodology, Supervision, Funding acquisition}
\author[c]{Danny~De~Vleeschauwer}
\credit{Writing - Review \& Editing, Conceptualization}
\author[a]{Yorick~De~Bock}
\credit{Conceptualization, Software}
\author[a]{Nina~Slamnik-Kriještorac}
\credit{Writing - Review \& Editing}
\author[c]{Chia-Yu~Chang}
\credit{Writing - Review \& Editing}
\author[b]{Natalia~Gaviria}
\credit{Writing - Review \& Editing, Conceptualization}
\author[a]{Erik~Mannens}
\credit{Writing - Review \& Editing}
\author[b]{Juan~F.~Botero}
\credit{Writing - Review \& Editing, Supervision, Conceptualization}
\author[a]{Steven~Latré}
\credit{Writing - Review \& Editing, Supervision}

\affiliation[a]{
            organization={University of Antwerp - imec, IDLab},
            addressline={Sint-Pietersvliet 7}, 
            city={Antwerp}, 
            postcode={2000},
            country={Belgium}}

\affiliation[b]{
            organization={Universidad de Antioquia},
            addressline={Calle 67 No. 53-108}, 
            city={Medellín}, 
            country={Colombia}}
\affiliation[c]{
            organization={Nokia Bell Labs},
            addressline={Copernicuslaan 50},
            city={Antwerp},       
            postcode={2018},
            country={Belgium}} 

\cortext[1]{Corresponding author}

\begin{abstract}
Automating network processes without human intervention is crucial for the complex \ac{6g} environment.  Thus, 6G networks must advance beyond basic automation, relying on \ac{ai} and \ac{ml} for self-optimizing and autonomous operation. This requires zero-touch management and orchestration, the integration of \ac{ni} into the network architecture, and the efficient lifecycle management of intelligent functions. Despite its potential, integrating \ac{ni} poses challenges in model development and application. To tackle those issues, this paper presents a novel methodology to manage the complete lifecycle of \ac{rl} applications in networking, thereby enhancing existing \ac{mlops} frameworks to accommodate \ac{rl}-specific tasks. We focus on scaling computing resources in service-based architectures, modeling the problem as a \ac{mdp}. Two \ac{rl} algorithms, guided by distinct \acp{rfn}, are proposed to autonomously determine the number of service replicas in dynamic environments.

Our proposed methodology is anchored on a dual approach: firstly, it evaluates the training performance of these algorithms under varying \acp{rfn}, and secondly, it validates their performance after being trained to discern the practical applicability in real-world settings. We show that, despite significant progress, the development stage of \ac{rl} techniques for networking applications, particularly in scaling scenarios, still leaves room for significant improvements. This study underscores the importance of ongoing research and development to enhance the practicality and resilience of \ac{rl} techniques in real-world networking environments.
\end{abstract}




\maketitle

\section{Introduction}
\glsresetall

The upcoming 6G networks will face challenging and diverse objectives, such as offering \ac{qos} while guaranteeing \ac{qoe}, infrastructure optimization, and scalable resource utilization. To face these challenges, these networks necessitate advanced automation beyond simplistic rules and heuristics~\cite{giordani2020toward}. Recognizing this complexity, \ac{ai} and \ac{ml} are acting as main enablers for network automation and, ultimately, for providing the network with self-X (where "X" can stand for healing, configuration, management, optimization, and adaptation) capabilities~\cite{taleb2023ai, coronado2022zero}. 

By embedding intelligence across all layers and throughout the entire lifecycle of communication services, the telco industry will transition towards an \ac{ai}-Native environment. The primary objective is to enable highly autonomous networks guided by high-level policies and rules. Emphasizing closed-loop operations and leveraging \ac{ai} and \ac{ml} techniques, the creation of \ac{ni} emerges as a pipeline of efficient algorithms for rapid response to network events~\cite{camelo2022requirements}, enabling autonomous network operations, and minimizing human intervention~\cite{etsi-zsm}. 

This transition will offer new opportunities for service provisioning but also introduce technical and business challenges. One notable innovation is the emergence of a \ac{nio} that supports the orchestration and management of such \ac{ai}/\ac{ml} models~\cite{camelo2022daemon}. In general, appropriate workflows for \ac{ni} lifecycle management should be provided~\cite{chatzieleftheriou2023orchestration}, which require the alignment with \ac{mlops} practices~\cite{zhang2023operationalizing}. The existing \ac{mlops} frameworks strive to automate and operationalize \ac{ml} processes, ensuring the delivery of production-ready software. These workflows are designed to be model- and platform-agnostic. However, most \ac{mlops} implementations primarily focus on \ac{sl}, often neglecting comprehensive support for \ac{rl} algorithms~\cite{li2022rlops}.

Therefore, to enable a completely autonomous network operation, the \ac{nio} should interpret the high-level policies and rules, e.g., using intent-based management~\cite{mehmood2023intent}, and guarantee that if needed, the \ac{ni} is ready for composing the \ac{ns} and posterior deployment. To achieve this, the \ac{nio} should trigger the training of the \ac{ai}/\ac{ml} models and select the models that best suit the network's operational conditions and requirements. With multiple \ac{ai}/\ac{ml} models available to address a networking problem, the challenge lies in determining which model to deploy and identifying the necessary metrics for this selection.


Focusing on the scaling computing resources using \ac{rl} in next-generation networks as a use case, this paper follows the \ac{mlops} workflow (cf. Figure~\ref{fig:mlops}) to describe how the \ac{nio} should tackle the challenges presented in each of the stages of the \ac{ni} lifecycle~\cite{li2022rlops}. Properly tackling these challenges, as we intend to do in this paper, will allow 6G systems to be closer to a completely autonomous network operation, which will improve the scalability, reliability, and performance of network services leveraged by \ac{ai}/\ac{ml} algorithms. The contributions of this paper are manifold.

\begin{itemize}
    \item We propose, as a use case, the scaling of computing resources in service-based network architectures such as \ac{oran} or \ac{nfv}. This use case requires intelligent scalers since current approaches for scaling are based on \ac{cpu} usage. However, under this approach, the correlation between \ac{sla}, \ac{qos} parameters, scaling decision triggers, and learning metrics is unknown or difficult to model. For instance, Altaf et al. and Gotin et al.~\cite {altaf2018auto, gotin2018investigating} show that \ac{cpu} utilization might not be the best metric for non-\ac{cpu} intensive applications. We model the scaling problem as an \ac{mdp} and propose two \ac{rl} algorithms that autonomously determine the number of replicas under a given constraint and constantly changing environment.
    
    \item We design three \acp{rfn} suitable for the scaling problem in multi-objective \ac{rl}. These \acp{rfn} guide agents to explore different strategies to achieve their objectives. Sparse \acp{rfn} provide rewards only after a sequence of actions, making learning challenging, while others offer frequent feedback, aiding faster learning and convergence. This setup helps explore a wide range of potential solutions, leading to more robust and adaptive learning. It also examines how agents balance competing objectives, crucial in multi-objective scenarios.

    \item We enhance the \ac{mlops} methodology by incorporating practices for algorithm selection and comparison. In contrast to prior methodologies such as \ac{mlops}~\cite{kreuzberger2023machine}, \ac{rlops}~\cite{li2022rlops}, Q-Model~\cite{kurrek2020q}, and \ac{oran} workflows~\cite{oran-wg2}, our approach trains, validates and compares  multiple \ac{rl} models configured with different \acp{rfn}. This adds complexity but aims to optimize across various algorithms. The methodology is validated through extensive experimentation and performance evaluation. 
    
    \item We propose a benchmark for testing and comparing scaling agents and their algorithms, built on established \ac{rl} frameworks like Stable Baselines~\cite{stable-baselines3} and OpenAI Gym~\cite{brockman2016openai}. This benchmark, open-sourced with baseline agents, facilitates the evaluation, comparison, and advancement of \ac{rl} algorithms. It complements existing gym environments for networking~\cite{gawlowicz2019ns}.
\end{itemize}

\begin{figure}
    \centering
    \includegraphics[width=0.9\textwidth]{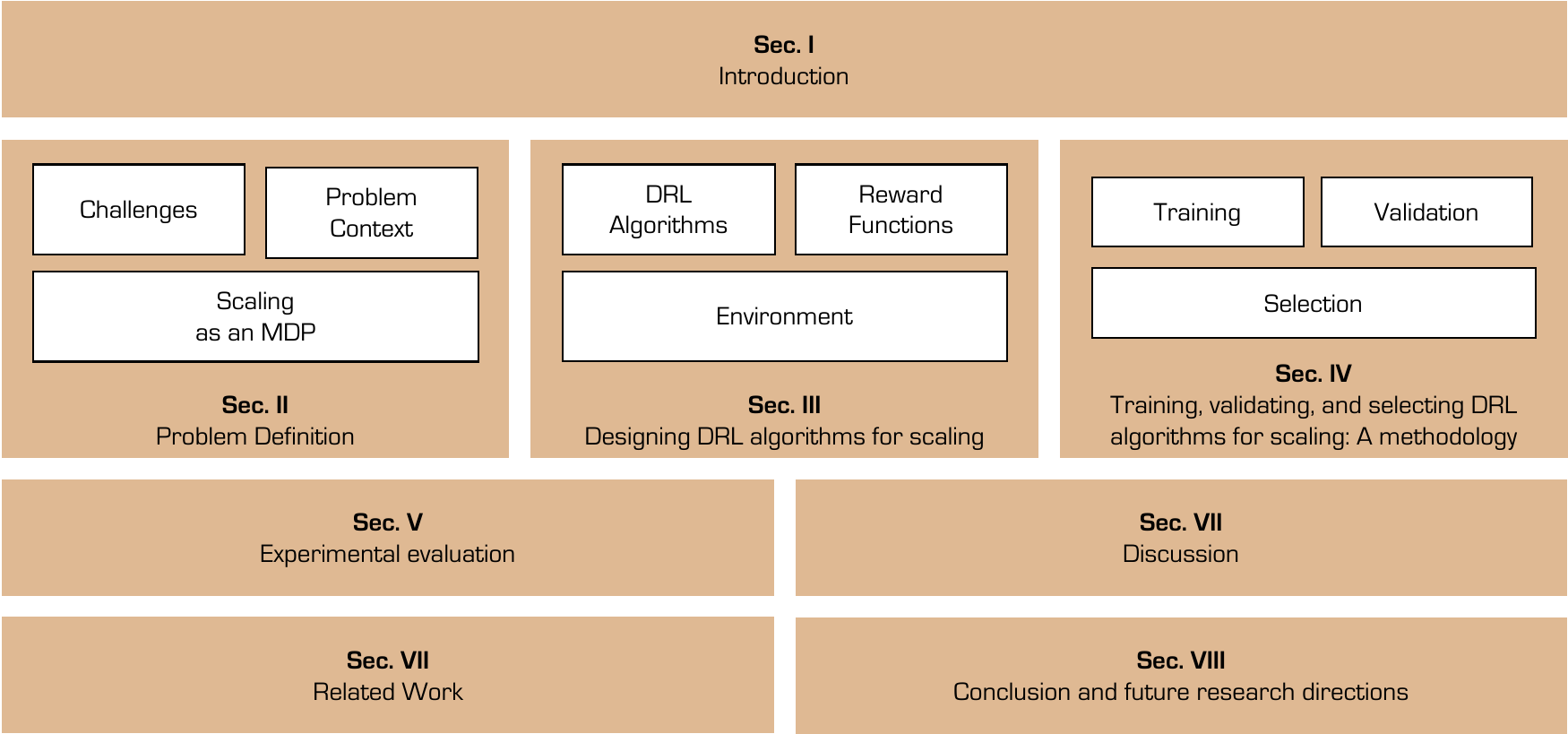}
    \caption{Overview of the organization of the paper.}
    \label{fig:organization}
\end{figure}

The remainder of this paper is organized as sketched in Figure~\ref{fig:organization}, which is divided into eight sections. This Section introduces the paper. Section~\ref{sec:problem} mentions the challenges in operationalizing \ac{ml} in networking, briefly explains scaling and how it can be modeled as an \ac{mdp}. In Section~\ref{sec:design}, we introduce two \ac{drl} algorithms to solve the \ac{mdp} and define three distinct \acp{rfn}, explaining how they were designed and the rationale behind each \ac{rfn}. Section~\ref{sec:develop-verify} defines the methodology we employed to benchmark \ac{rl} algorithms for scaling, while Section~\ref{sec:results} details the experiments we conducted to test and validate the methodology for model training, validation, and selection; in Section~\ref{sec:discussion}, we discuss the results. Finally, Section~\ref{sec:sota} reviews some state-of-the-art approaches for scaling and \ac{rl} benchmarking and Section~\ref{sec:conclusion} concludes the paper.
\section{Problem definition} \label{sec:problem}  

This section outlines the problem we aim to solve from two distinct perspectives. Firstly, we assume that the development and operation of \ac{ni} should be conducted at the \ac{nio} level. We emphasize the necessity of a methodology for evaluating and selecting NI algorithms, especially the ones based on \ac{rl}. Additionally, this section addresses the definition of the problem, focusing on modeling the dynamic scaling of computing resources in service-based network architectures. To facilitate this, we provide contextual background information to comprehend the scaling problem and its implications in the design of \ac{rl} algorithms.

\subsection{Challenges in operationalizing RL-based algorithms}\label{sec:motivation}



\begin{figure}
    \centering
    \includegraphics[width=0.7\textwidth]{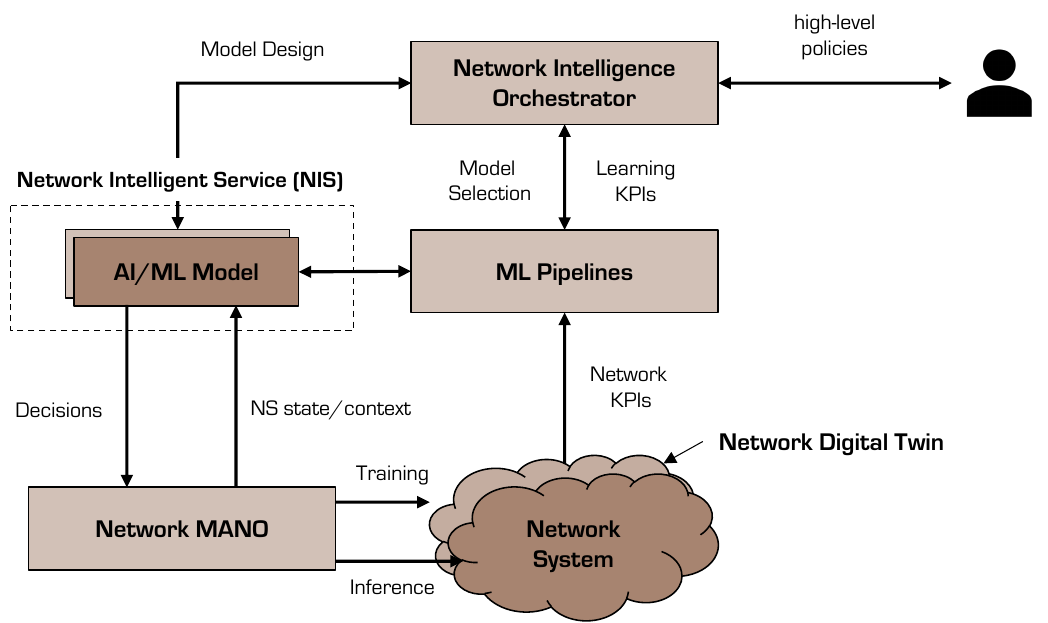}
    \caption{Developing and operating NI in AI-native architectures requires interaction among different building blocks; among them, the NI Orchestrator plays a fundamental role in coordinating the remaining elements~\cite{camelo2022daemon}. 
    }
    \label{fig:mlops}
\end{figure}

For developing and operating \ac{nis}, an effective \ac{ml} workflow should be applied alongside the \ac{nis} lifecycle, aligned with \ac{mlops} practices~\cite{zhang2023operationalizing}. A key component in an \ac{ai}-native architecture with autonomous operations is the \ac{nio}. This orchestrator is responsible for interpreting the user intentions (e.g., based on intent-based networking~\cite{mehmood2023intent}), composing a \ac{nis}, deploying an appropriate \ac{ai}/\ac{ml} model, which guides the \ac{nis} behavior, and monitoring that the \ac{nis} behavior is as expected. To be able to do that, the \ac{nio} implements (on its own or via a third-party service) an \ac{ml} workflow. Figure~\ref{fig:mlops} illustrates this closed-loop system. 

This workflow starts with defining the requirements, typically derived from the problem at hand. This phase involves the selection of an appropriate learning type and \ac{ml} model, as well as the preparation of the data on which the model is trained on. For instance, scaling can be solved through \ac{sl} techniques in the form of time series or by forecasting the incoming workload, both of them require the collection, curation, and maintenance of a database with historical workload values and scaling decisions and values of the \acp{kpi} collected from the network. However, scaling can also be solved through \ac{rl} techniques (as it will shown later in this paper), which implies the preparation of a sandbox~\cite{wilhelmi2021usage} in which the \ac{rl} agent can freely interact with a digital replica~\cite{almasan2022network} of the network to train a scaling policy.  


Once the \ac{ml} model is selected, the model training and fine-tuning occur thanks to the \ac{mlops} framework, including tuning the model's architecture and hyperparameters (e.g., learning rate, activation functions, and regularization methods). Following model construction, validation is performed by introducing unseen data to enhance generalization. Upon achieving a predefined performance, the \ac{nio} deploys the model as a \ac{nif} inside the \ac{nis}. After deployment, the \ac{nio} continuously monitors the performance of the \ac{nis}. Typical quality metrics of \ac{ml} models are accuracy, loss value, and average return after an epoch, among others. Nevertheless, the \ac{nio} could potentially be supported by metrics gathered from the network to assess the quality of the \ac{ml} model. Continuous monitoring is vital due to potential data and model drift, ensuring service quality. If the model fails to meet quality standards, refinement is undertaken, restarting the cycle.

The existing \ac{mlops} frameworks strive to automate and operationalize \ac{ml} processes, ensuring the delivery of production-ready software. The workflow is intended to be model- and platform-agnostic. However, most of the \ac{mlops} implementations predominantly focus on \ac{sl}, often neglecting comprehensive support for \ac{rl} algorithms~\cite{li2022rlops}. Incorporating \ac{rl} introduces additional challenges to the \ac{ml} workflow, encompassing four aspects.

Firstly, the problem to be addressed must be formulated as a \ac{mdp}, where the learning goal of a \ac{rl} agent is to find an optimal policy through the use of a \ac{rfn}. By interacting with an environment, the \ac{rl} agent gathers states and rewards, facilitating the iterative approximation or computation of the expected long-term reward, enabling the selection of optimal actions at each step. Unlike the intrinsic ``labels'' in \ac{sl}, rewards reflect the anticipated behavior of agents. Consequently, the design of the \ac{rfn} must be done carefully and sometimes requires specialized considerations. 

Interactions and learning for \ac{rl} algorithms still come along with significant challenges due to network vulnerabilities. To ensure the safety of \ac{rl} algorithm training, an offline approach is recommended~\cite{oran-wg2}. In this offline setting, the \ac{rl} agent accumulates experiences by exploring actions randomly, allowing it to learn without impacting network operations. To facilitate this learning process, \acp{ndt} are proposed to provide a controlled, reliable, and easily accessible simulation environment~\cite{almasan2022network}, enabling \ac{rl} algorithms to explore new actions safely. Consequently, a parallel branch is required to support this learning type. In the primary branch, decisions from an already trained agent are implemented in the production network, while algorithms still undergoing training can introduce their outcomes in the \ac{ndt}~\cite{camelo2022daemon}.

A third aspect challenging the full support of \ac{rl} in \ac{mlops} frameworks is the reproducibility issue. It has been found that minor implementation details can considerably impact performance, sometimes surpassing the disparity between various algorithms~\cite{engstrom2019implementation, islam2017reproducibility, henderson2018deep}. Best practices in the \ac{rl} community include the use of standard agent-environment interfaces~\cite{brockman2016openai}, \ac{rl} implementations~\cite{stable-baselines3} and full transparency in reporting the settings used when training \ac{rl} algorithms. Unfortunately, such practices are not fully adopted in the growing body of literature of \ac{rl} applications in networking~\cite{luong2019applications}. 

Finally, a key feature of the \ac{nio} involves autonomously selecting and validating models. In a dynamic environment where various \acp{nsp} may develop their own \ac{ai}/\ac{ml}/\ac{rl} models, multiple models could potentially be deployed for a single function. Hence, the orchestrator is crucial in determining which available model should be deployed, preventing conflicts, and optimizing resource utilization. Unfortunately, existing implementations of \ac{mlops} frameworks lack this feature, as their workflows are designed for a single model or perform manual model selection based solely on the learning metrics (e.g., loss, accuracy, or reward). However, as we will show in the following sections, \ac{rl} algorithms should also be validated in terms of the performance of the task they are designed to solve. 

\subsection{Scaling in service-based network architectures} \label{sec:system_model}
The evolution of current and future networks is characterized by a shift towards a service-oriented architecture, where essential \acp{nf} are increasingly implemented as modular software components. These components are often encapsulated within lightweight containers that can operate on \ac{cots} hardware~\cite{akyildiz20206g}. The \ac{oran} architecture exemplifies this trend, as it disaggregates and virtualizes \ac{ran} deployments, turning them into software-based entities with clearly defined interfaces that facilitate interoperability across different vendors~\cite{polese2023understanding}.

In this context, softwarized \acp{nf}, commonly referred to as \acp{vnf}, can be orchestrated to form a \ac{ns}. This granular decoupling offers numerous advantages, including enhanced automation, streamlined system integration, and efficient workflows with shared resources. These resources can be programmed to optimize various performance and cost objectives. By decoupling hardware and software, \acp{nsp} can lower equipment costs and energy consumption, as a single platform can be utilized for multiple applications, users, and tenants. Combined with \ac{sdn}, these service-based network architectures enable more flexible and dynamic network management~\cite{bonfim2019integrated}. \acp{sla} play a crucial role in formalizing stakeholder relationships, specifying \ac{ns} requirements such as the maximum latency a \ac{ns} can tolerate. Consequently, \acp{nsp} can adjust the size of their \acp{ns} to meet user requirements while reducing \ac{capex} and \ac{opex} through dynamic resource provisioning~\cite{bonfim2019integrated}.  

One of the key operations in this context is scaling, which involves adjusting the resources allocated to a \ac{ns} based on demand~\cite{etsi-nfv-mano}. The goal is to minimize resource allocation during periods of low demand, thereby reducing costs while ensuring sufficient resources are available during peak periods~\cite{adamuz2018automated}. Achieving this balance requires careful consideration of the cost of deploying multiple \ac{vnf} replicas and adhering to the quality objectives outlined in the \ac{sla}. Typically, scaling decisions are informed by monitoring infrastructure performance metrics, such as \ac{cpu}, memory, and storage usage, and by defining operational thresholds that trigger the addition or removal of resources. 

Scaling involves the interaction of three dynamic processes over the same infrastructure. First, the workload $j_t$ represents the number of jobs to be processed by the \ac{ns} at time $t$, such as the number of users served by the \ac{ns}. Second, the latency experienced by a job $d_t$, defined as the time to process that job plus any waiting time, depends on the availability of \acp{vnf}. More \acp{vnf} reduce the processing time, but some latency is inevitable due to the random distribution of jobs within a time slot. Finally, the scaling algorithm must anticipate the number of \acp{vnf} $v_t$ required to process the workload so that $d_t$ remains within the limits specified in the \ac{sla}.

\subsection{Scaling as an MDP} \label{sec:mdp}

\begin{figure}
    \centering
    \includegraphics[width=0.6\linewidth]{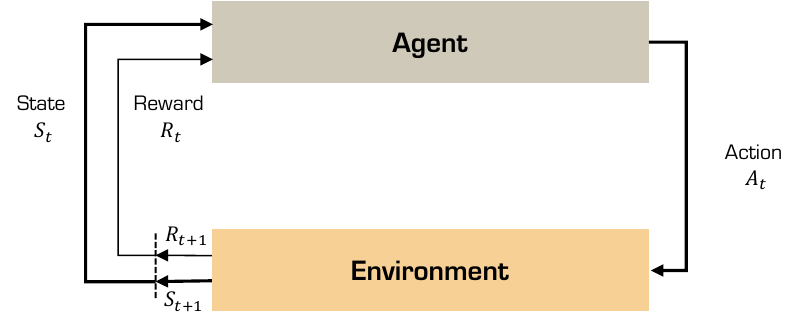}
    \caption{Basic interactions in a Markov Decision Process~\cite{sutton2018reinforcement}.}
    \label{fig:mdp-setup}
\end{figure}

An \ac{mdp} is a discrete-time stochastic framework that models sequential decision-making problems~\cite{sutton2018reinforcement}, where both the future state and the rewards depend only on the immediate state and actions. The basic entities in an \ac{mdp} are the agent and the environment. The agent is the entity that decides which action to take. Everything else the agent interacts with is called the environment. Such interactions are depicted in Figure~\ref{fig:mdp-setup}. Agent and environment interact in discrete time slots, $t = 1, 2, \ldots$. At each slot, the agent receives a representation from the environment's state $S_t \in \mathcal{S}$, and based on that, selects an action $A_t \in \mathcal{A}(s)$. In the next slot, $t+1$, the agent receives a reward $R_{t+1} \in \mathcal{R} \subset \mathbb{R}$ and the environment arrives to a new state, $S_{t+1}$. 

Mathematically, an \ac{mdp} is defined by the tuple $\langle \mathcal{S},\mathcal{A},P,r \rangle$ where $\mathcal{S}$ and $\mathcal{A}$ are the state and action spaces; $P: \mathcal{S} \times \mathcal{A} \times \mathcal{S} \rightarrow [0, 1]$ is the transition kernel, with $p(s'|s,a)$ denoting the probability of transitioning to state $s'$ from $s$ after action $a$ is taken. In an \ac{mdp}, the probability $p$ completely characterizes the dynamics of the environment. 

The optimization objective in an \ac{mdp} is to find a policy, a strategy for choosing actions in various situations, that maximizes cumulative rewards over time. For that $r: \mathcal{S} \times \mathcal{A} \times \mathcal{S} \rightarrow \mathrm{R}$ is the immediate reward the agent obtains for performing action $a$. The policy $\pi: \mathcal{S} \rightarrow \mathcal{P}(\mathcal{A})$, is a mapping function from the state space to the space of probability distributions over the actions, with $\pi(a|s)$ denoting the probability of selecting action $a$ in state $s$. The optimal policy maximizes the expected long-term return, where the return is defined as some specific function of the reward sequence. One common function is the discounted sum of immediate rewards, as shown in Equation~\ref{eq:discounted_return}, where $G_t$ is the expected discounted return and $0 \leq \gamma \leq 1$ is the discount rate. The notations used throughout this paper are summarized in Table~\ref{tab:notations}.

\begin{equation}
    G_t = \sum_{k=0}^{\infty} \gamma^k R_{t+k+1}
    \label{eq:discounted_return}
\end{equation}

\begin{table}[width=.9\linewidth,cols=2,pos=ht]
\caption{Variables used in this paper.}
\label{tab:notations}
\begin{tabular*}{\tblwidth}{@{} LL@{} } 
\toprule
\textbf{Variables} & \textbf{\textbf{Description}}                                \\ 
\midrule
$\mathcal{S}$      & State space.                                                 \\
$s, S_t$           & A particular state in step $t$, random variable.             \\ 
$\mathcal{A}$      & Action space.                                                \\ 
$a, A_t$           & A particular action in step $t$, random variable.            \\ 
$P$                & Transition Kernel.                                           \\
$p(s'|s,a)$        & Transition probability to state $s'$ by taking action $a$ in state $s$ .\\ 
$\pi(a|s)$         & Probability of selecting action $a$ in state $s$.            \\ 
$\mathcal{R}$      & Reward space.                                                \\
$r,  R_t$          & Immediate reward, random variable.                           \\
$G(t)$             & Expected discounted reward.                                  \\
$\epsilon$         & Tolerance range.                                             \\ 
$j_t$              & Number of jobs to be processed at step $t$,                  \\ 
$v_t$              & Number of active replicas in step $t$.                       \\ 
$\bar{v}$          & Mean number of active replicas during a period of time.      \\
$\bar{c}_t$        & Mean CPU usage of the active replicas in step $t$.           \\ 
$c_{tgt}$          & CPU utilization target.                                      \\ 
$d_t$              & Processing latency of the active replicas in step $t$.       \\ 
$\bar{d}$          & Mean latency of the active replicas during a period of time. \\ 
$d_{tgt}$          & The target latency. The maximum tolerated value $(1+\epsilon) \cdot d_{tgt}$ cannot be surpassed. \\ 
$c_{perf}$         & Performance cost.                                            \\ 
$w_{perf}$         & Importance (weight) of $c_{perf}$ in the total cost.         \\ 
$c_{res}$          & Resource cost.                                               \\ 
$w_{res}$          & Importance (weight) of $c_{res}$ in the total cost.          \\ 
$V'$               & Normalized number of replicas created by an agent.           \\ 
$w_{v}$            & Importance (weight) of $V'$ in networking scoring.           \\ 
$D'$               & Normalized latency of an agent.                              \\ 
$w_{d}$            & Importance (weight) of $D'$ in networking scoring.           \\
\bottomrule
\end{tabular*}%
\end{table}

Modeling scaling as a \ac{mdp} allows us to design the scaling solution as a closed-loop control, which leverages autonomicity and adaptiveness. 
A closed-loop control involves establishing a feedback system where the scaling of resources is continuously adjusted based on observed performance metrics. The loop starts by collecting 
data from the resources through distributed sensors, constituting the state $s$ of the system. For instance, the system can monitor factors like server load, response times, and resource utilization in cloud computing. Then, 
the received information is correlated, and a system model is created. This model allows the closed-loop control system in the 
decision-maker to automatically trigger action $a$ when deviations from optimal performance are detected, such as provisioning additional resources to meet demand or scaling down to save costs during periods of lower activity. Finally, 
the planned actions are executed over the resources. 



In this paper, we examine a scenario where a \ac{nsp} must deploy a service consisting of a single containerized \ac{vnf} (e.g., a video transcoder). Users or other services within the network can request this service by submitting processing tasks or jobs. To meet \acp{sla}, the \ac{nsp} resizes its \ac{ns} by adjusting the number of \ac{vnf} replicas to achieve both operational and business objectives in a multi-objective framework~\cite{adamuz2018automated}. We focus on horizontal scaling, which distributes the workload across multiple instances of the same application, as opposed to vertical scaling, which is limited by the capacity of a single server. 

\begin{figure}
    \centering
    \includegraphics[width=0.65\textwidth]{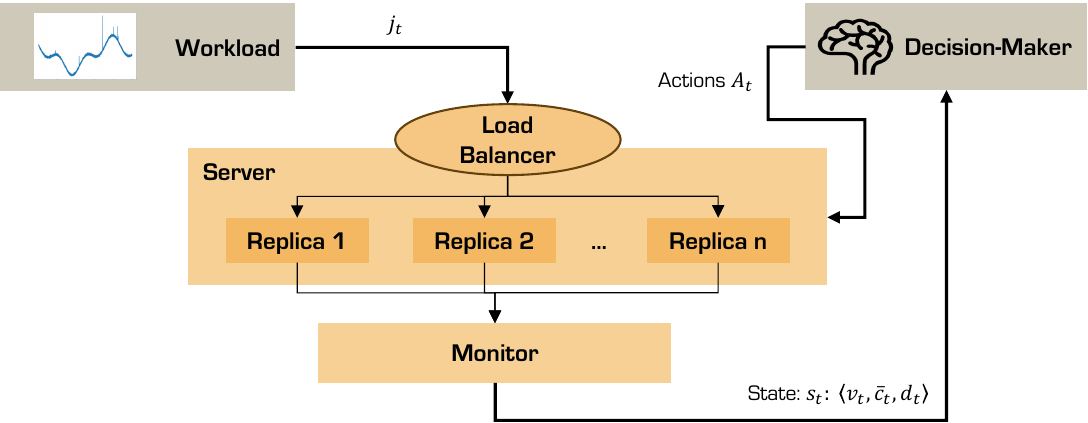}
    \caption{Simple model of the scaling problem.}
    \label{fig:system_model}
\end{figure}

The system operates in discrete time slots, $t = 1, 2, \ldots$. At the beginning of each slot, the workload is queued before a load balancer. The workload is evenly distributed among the active replicas if no jobs are in the queue. However, a job may still experience some latency even without a queue, as jobs are processed sequentially at the replica level. If a queue is present, incoming jobs must wait until an available replica can process them. Each replica can handle a predetermined number of jobs per time slot. The scaling algorithm then determines the number of \ac{vnf} replicas needed for the next slot based on the accumulated workload and expected future demand. However, neither the accumulated workload nor expected future demand is assumed to be known nor modeled. This process is repeated in each subsequent time slot. Figure~\ref{fig:system_model} illustrates the described system.

At time step $t$, the state $s_t$ is defined as a tuple $\langle v_t, \bar{c}_t, d_t\rangle$, where $v_t$ represents the number of active replicas, $\bar{c}_t$, is the mean \ac{cpu} usage among the active replicas, and $d_t$ is the time to process the job plus the waiting time in the queue, i.e., the peak latency introduced by processing the jobs. Based on this information, the decision-maker decides its actions. This simple model keeps the action space small by only defining three actions. Thus, $A_t$ is a number from the set $\mathcal{A} = \{-1, 0, 1\}$, where $-1$ means to decrease the number of replicas by one, $1$ means to increase the number of replicas by one, and $0$ will maintain the same number of replicas. Only one action is allowed per time slot. 

In the following sections, we focus on the design and development of \ac{rl}-based scaling algorithms, paying special attention to the challenges we just mentioned when applied to the networking field. In particular, this section poses the scaling problem as a \ac{mdp}, setting the environment and algorithm design requirements, shown in Section~\ref{sec:design}. Section~\ref{sec:develop-verify} focuses on the development part as it explains the difficulties when training and validating \ac{rl} algorithms. Additionally, it proposes a methodology for performing model selection, an important step before its deployment in production networks.
\section{Designing DRL algorithms for scaling} \label{sec:design}
This section explores the intricacies of designing \ac{drl} algorithms tailored explicitly for efficiently scaling computing resources. Based on the problem constraints elaborated in Section~\ref{sec:problem}, we discuss the nuanced aspects of algorithmic design, elucidating key principles and methodologies essential for achieving proper lifecycle management of this type of \ac{ni} in service-based architectures. 

\subsection{DRL algorithms tailored to the Scaling problem}\label{sec:design:algorithms}
\ac{rl} has become a key approach for solving \acp{mdp} due to its ability to explore and evaluate different actions within an environment, making it ideal for dynamic and uncertain situations. Its strength lies in adapting to various environments and balancing exploration with exploitation to find optimal policies, making it a highly effective tool for solving \acp{mdp}~\cite{sutton2018reinforcement}.

Almost all \ac{rl} algorithms estimate how good it is for the agent to perform a given action in a given state. The value function of a state, $v_\pi(s)=\mathbbm{E}_\pi[G_t | S_t=s], \text{for all } s \in \mathcal{S}$, and the action-value function, $q_\pi(s,a)=\mathbbm{E}_\pi[G_t | S_t=s, A_t=a]$, are the expected return when starting in $s$ and following $\pi$, and taking action $a$, respectively, where $\mathbbm{E}_\pi[\cdot]$ denotes the expected value of a random variable. Thus, as the objective in an \ac{rl} setup is to maximize the expected return, $G_t$, the agent should update its policy $\pi$ to select the actions that lead them towards that goal by estimating the value function $v_\pi(s)$ or the action-value $q_\pi(s,a)$. Such value functions can be estimated from experience.

An optimal policy, $\pi_*$, is defined as the policy whose expected return is greater than or equal to that of any other policy $\pi' \: \forall s \in \mathcal{S}$. A higher state-value function characterizes such a policy, $v_*(s)=\underset{\pi}{\max} \; v_\pi(s), \; \forall s \in \mathcal{S}$, where $v_*(s)$ denotes the optimal state-value function with the highest value. Optimal policies also share the same optimal action-value function, $q_*(s,a) = \underset{\pi}\max \; q_\pi(s,a)  \; \forall s \in \mathcal{S}$. Then, if the optimal value functions can be found, the optimal policy can be obtained using Equation~\ref{eq:optimal_policy}. This is the working principle of many \ac{rl} algorithms.

\begin{equation}
   \pi_*(s) = \underset{a \in \mathcal{A}}{\arg\max} \; q_*(s,a)
   \label{eq:optimal_policy}
\end{equation}

When the state and action spaces derived from the \ac{mdp} are sufficiently small, the value functions of each state-action pair are stored as arrays or tables~\cite{sutton2018reinforcement}. In this case, the approaches can find exact solutions; that is, they can find the optimal value function and the optimal policy if they visit each state-action pair an infinite number of times. 

However, if the state and action spaces are sufficiently big, the optimal value function is approximated. Such is the case of scaling. Since the data composing the state is typically real-valued, a tabular approach lacks scalability due to the quantization of the monitored samples, making this approach impractical for scaling as the size of the tables is unpractically large. On the contrary, \ac{drl} is preferred for complex, high-dimensional, or continuous state and action spaces. Under \ac{drl}, a deep neural network approximates the value function, policy, or both. These neural networks can handle high-dimensional and uncountable infinite state spaces.

\ac{drl} algorithms can be classified as on-policy or off-policy. Off-policy algorithms update their policy using data generated by a different policy. This means that the agent can learn from past experiences and use data collected by any previous policy, making them more sample-efficient but potentially introducing higher variance in the learning process. On the other hand, on-policy algorithms update their policy while using the data generated by the same policy. This means that the data collected during the learning process is directly used to improve the current policy, leading to more stable but potentially less sample-efficient performance. As a result, the policy being learned is constantly changing as the agent explores the environment and gathers new experiences. Other classifications of \ac{drl} techniques are possible. We refer the reader to~\cite{arulkumaran2017deep} for a detailed view of the state-of-the-art of these kinds of algorithms. 

Given the properties of the scaling problem, there are two known \ac{drl} algorithms that are suitable, namely, a \ac{dqn}~\cite{mnih2015human} and \ac{ppo}~\cite{schulman2017proximal} algorithm. \ac{dqn} is an off-policy, value-based algorithm that combines Q-learning with deep neural networks to estimate the value of states or state-action pairs. Being a value-based algorithm, \ac{dqn} learns the value of each action in a given state to later select the actions that maximize the expected reward over time. Through repeated interactions and learning from the experience replay buffer, the \ac{dqn} can learn an optimal policy to perform well in complex tasks. Still, it may struggle with high-dimensional state spaces. It has been successfully applied to various tasks, including playing Atari games and controlling robotic systems~\cite{mnih2013playing}. 

On the other hand, \ac{ppo} is an on-policy, policy-based algorithm based on policy gradient. Policy-based \ac{rl} directly parameterizes the policy mapping states to actions without explicitly learning a value function, offering greater flexibility and robustness in exploring the action space. However, more samples may be required to converge to an optimal policy than value-based methods. \ac{ppo} tries to address the limitations of previous policy gradient algorithms such as \ac{trpo} by striking a balance between making significant updates to the policy (to improve performance) and ensuring that the updates are not too large, which could lead to instability or catastrophic changes. \ac{ppo} achieves this by updating the policy with a ``clipped'' objective function. 

\subsection{Reward Functions (RFns)} \label{sec:design:rfs}
We designed three \acp{rfn} to guide the agents' learning. The goal in proposing multiple \acp{rfn} is twofold. On one side, we want to highlight that the \ac{rfn} definition is among the most difficult aspects in \ac{rl}. Here, we have three different \acp{rfn} for the same problem, and though they seem to be logical and make sense, not all of them yield good results, as shown in Section~\ref{sec:results}. Conversely, the exploratory work performed in this study may help uncover a broader spectrum of potential solutions and can lead to more robust and adaptive learning.

\subsubsection{Reward Function 1 (RFn1): Inspired from cart-pole}
Contrary to most of the \ac{rl} applications in networking, where the states, actions, and \ac{rfn} are defined using a networking rationale, in this \ac{rfn}, we map the auto-scaling problem to known applications of \ac{rl}. One of the classical environments in OpenAI Gym is \textit{Cart-Pole},\footnote{\url{https://www.gymlibrary.dev/environments/classic_control/cart_pole/}} where a pole is attached to a cart moving along a frictionless track. The pole is placed upright on the cart, and the goal is to balance it by applying forces to the left and right sides of the cart. Similarly, in scaling, the agent tries to guarantee a given \ac{sla} (e.g., latency) by taking discrete actions (i.e., increase, decrease, or maintain the number of replicas) that resemble the ones taken in \textit{Cart-Pole} by applying forces to the left or the right. Consequently, the \ac{rfn} is defined similarly as in the \textit{Cart-Pole} problem. More specifically, the \ac{rfn} at step $t$ is defined as

\begin{equation}
    r_t = 
    \begin{cases}
        1 & |d_t - d_{tgt}| < \epsilon \cdot d_{tgt} \quad \lor\\
         & |\bar{c}_t - c_{tgt}| < \epsilon \cdot c_{tgt} \\
        0 & |d_t - d_{tgt}| \geq \epsilon \cdot d_{tgt} \quad  \lor \\
         & |\bar{c}_t - c_{tgt}| \geq \epsilon \cdot c_{tgt}  \\
        -100 & \text{if wrong behavior}
    \end{cases}
    \label{eq:reward1}
\end{equation}

where $d_{tgt}$ is the target latency as defined by the \ac{sla} and $\epsilon$ is a range of tolerance (e.g., \num{20}\%). If the \ac{rfn} is only defined based on the perceived latency, the agent will take the most obvious action: to keep increasing the number of instances, disregarding the economic impact of such a decision. To keep the number of instances at an adequate level, we also reward the agent if the current mean \ac{cpu} usage, $\bar{c}_t$, is within a predefined range. If the mean \ac{cpu} usage is low, probably the workload can be served using fewer replicas and vice versa. Moreover, the agent is severely penalized if it incurs in a wrong behavior, such as creating more replicas than necessary or exceeding the perceived latency beyond an allowed upper bound. Notice that the \ac{cpu} utilization target, $c_{tgt}$, is not normally specified by the \ac{sla} and must be approximated.  

\subsubsection{Reward Function 2 (RFn2): A Markov Reward Process}
A \ac{mrp} is a mathematical framework used to model and study the behavior of a system that involves stochastic transitions between states and yields rewards over time. \acp{mrp} are used to formalize problems where an agent interacts with an environment, and the agent receives rewards for being in certain states. This way, \acp{mrp} could be directly applied to \ac{rl} setups. A \ac{mrp} is defined by mainly three concepts: the states, the transitions, and the rewards. Notice that it is not necessarily the case that the states of the \ac{mrp} are the same as in the \ac{mdp}. 

Particularly for the scaling problem, we define two states: the process exhibits a latency above the specified threshold, or the process shows a latency below the defined threshold. The agent dictates the transitions between these two states with its actions. For example, it can be that the process is below the latency threshold, but the agent decides to remove a replica, which takes the process to the state of being above the latency threshold. Then, the algorithm designer can establish what actions led to a good outcome and which did not. 

\begin{figure}
    \centering
    \includegraphics[width=0.7\textwidth]{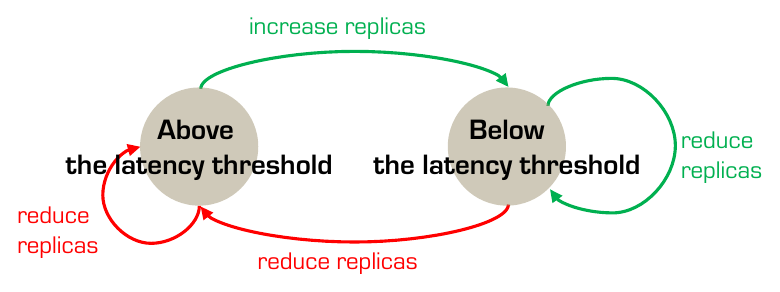}
    \caption{Markov Reward Process for auto-scaling}
    \label{fig:rf2_definition}
\end{figure}

Figure~\ref{fig:rf2_definition} shows an example of such \ac{mrp}. In green are depicted the actions that we consider good because they lead toward the goal we want to achieve, i.e., fulfilling the \ac{sla} with the least amount of replicas possible. In contrast, depicted in red are the actions that push us away from that goal. More specifically, for both agents, we only reward the good actions. We noticed that by using positive reinforcement, the agents could show a better performance. Then, the \ac{rfn} we used was defined as follows. 

\begin{itemize}
    \item If the process was in state \textit{Above}, and the agent's action was to increase the number of replicas, which led to the process being in the \textit{Below} state, the agent receives a reward of $1$. In this case, the agent is encouraged to increase the number of replicas only to fulfill the \ac{sla}.
    \item If the process was in state \textit{Below}, and the agent's action was to decrease the number of replicas, which led the process to keep being in the \textit{Below} state, the agent receives a reward of $1$. In this case, the agent is encouraged to use resources efficiently when the \ac{sla} is fulfilled.
    \item All the other combinations of states and transitions are not rewarded nor penalized.
\end{itemize}

\subsubsection{Reward Function 3 (RFn3): A Multi-Objective Function}
This \ac{rfn} aims to fulfill the agreed \ac{sla} with the minimum amount of replicas. Therefore, in every time step, the agent pays an immediate cost depending on how good or bad the action it took is. The cost of taking action when the environment moves from one state to another can be defined as a weighted function, including the following contributions. 

\begin{itemize}
    \item If the agent cannot fulfill the \ac{sla}, it incurs a performance cost $c_{perf}$, with an associated $w_{perf}$, which is paid every time the perceived latency $d$ exceeds a maximum tolerated latency, $(1+\epsilon) \cdot d_{tgt}$. The cost is zero otherwise. 
    \item If the agent must deploy a new replica, a resource cost $c_{res}$ is paid, with an associated $w_{res}$; this can be seen as a rental cost in cloud environments or the consumed energy of the replica while it is running.
\end{itemize}

These two contributions are combined into a weighted function, shown in Equation~\ref{eq:reward3:total-reward}, where the respective non-negative weights define an optimization profile, $w_{perf}+w_{res}=1$. The weights ($w_{perf}$ and $w_{res}$) multiply an indicator function $\mathbbm{1}\{\cdot\}$ that varies between \num{1} and \num{-1} depending on whether or not a condition is met, as indicated in Equations~\ref{eq:reward3:perf-ind} and~\ref{eq:reward3:res-ind}. For instance, if the perceived latency exceeds a maximum tolerated value, the indicator function is \num{1}; otherwise, the indicator function is \num{0}. Conversely, if a new replica is instantiated, the indicator function is \num{1}, or \num{-1} if the replica is removed. Finally, the \ac{rfn} is defined as the negative of the cost function.

\begin{align} 
r=-c_{total}= -(c_{perf} + c_{res}) \label{eq:reward3:total-reward}\\
c_{perf} = w_{perf} \cdot \mathbbm{1}\{perf\} \label{eq:reward3:perf-cost} \\
c_{res} = w_{res} \cdot \mathbbm{1}\{res\}  \label{eq:reward3:res-cost}\\
\mathbbm{1}\{perf\} = 
    \begin{cases} 
        1 & d_t > (1+\epsilon) \cdot d_{tgt}\\
        0 & \text{otherwise} 
    \end{cases}  \label{eq:reward3:perf-ind} \\
\mathbbm{1}\{res\}= 
    \begin{cases} 
        -1 & a = \text{remove replica}\\
        0 & a = \text{maintain replica} \\
        1 & a = \text{add replica}  
    \end{cases}
    \label{eq:reward3:res-ind}
\end{align}

Similar to our previous work~\cite{soto2023network}, we used different optimization profiles for this \ac{rfn} by assigning weights. By giving equal weights, the agent should learn policies that balance the creation of replicas while not exceeding the latency threshold. On the other hand, by assigning more weight to the resource cost, the agent should learn policies that limit the creation of replicas, disregarding the impact on latency violation. Conversely, if more weight is assigned to the performance cost, the agent should learn policies that encourage the creation of replicas to minimize the latency violation. Table~\ref{tab:opt-profiles-rf3} shows the optimization profile we used in this study.   

\begin{table}[width=.5\linewidth,cols=3,pos=ht]
    \caption{Optimization profiles used in RFn3.}
    \label{tab:opt-profiles-rf3}
    \begin{tabular*}{\tblwidth}{@{} CCC@{} }
    \toprule
    \textbf{\begin{tabular}[c]{@{}c@{}}Optimization\\ Profile\end{tabular}} & \textbf{\textbf{$w_{perf}$}} & \textbf{$w_{res}$} \\
    \midrule
    1 & 0.5  & 0.5  \\ 
    2 & 0.01 & 0.99 \\ 
    3 & 0.99 & 0.01 \\ 
    \bottomrule
    \end{tabular*}
\end{table}

\subsection{Environment} \label{sec:design:environment}

In light of the growing amount of research involving \ac{rl} algorithms in networking, creating a benchmarking environment is essential to assessing and comparing the performance of the different algorithms. To address this, we created a dedicated environment to facilitate fair and comprehensive assessments of \ac{rl} algorithms for scaling. This environment is tailored to the specific challenges and characteristics of the scaling problem, with the primary objectives of providing a common ground for evaluation, promoting collaboration within the research community, and tracking the progress of \ac{rl} algorithms over time. In this endeavor, we consider essential components such as the definition of different \acp{rfn}, the choice of evaluation metrics, and the design of test scenarios, all of which will contribute to the effectiveness of our benchmarking framework.

One of the main components in an \ac{mdp} is the environment. An environment is a process that interacts with the agent and reacts to the agent's actions by transitioning from one state to another. In the case of scaling, the environment is the system that hosts the different replicas and allows them to process the workload. 

For this evaluation, we developed \textit{DynamicSim}, a simulator that enables the creation of edge-cloud network scenarios. This simulator is based on \ac{Sim-Diasca}\footnote{\url{https://olivier-boudeville-edf.github.io/Sim-Diasca/}}, a general-purpose, parallel, and distributed discrete-time simulation engine for complex systems written in the Erlang language~\cite{song2011performance}. Erlang facilitates the implementation of large-scale parallel and distributed applications such as the ones tackled in networking. Erlang is a functional programming language based on the actor model, a powerful model for creating highly concurrent, distributed software. 

Each actor in this architecture is considered a processing unit, which can communicate with other actors via asynchronous messages. Each actor stores their respective messages that are pending processing. After processing the message, an actor modifies its state, sends more messages, or makes new actors. Due to the actors' lack of shared state or resources and asynchronous communication, this paradigm significantly reduces blocking waits and race situations, two major issues with concurrent systems. Moreover, the actor model facilitates distributed software development since it makes no difference if two actors are running on the same machine or different ones. 

\begin{figure}
    \centering
    \includegraphics[width=0.8\columnwidth]{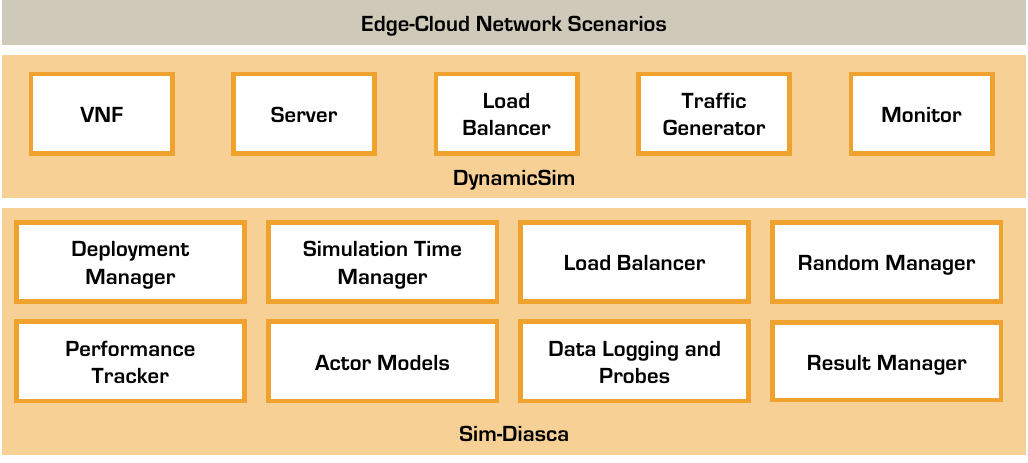}
    \caption{Architecture of \ac{Sim-Diasca}~\cite{song2011performance}}
    \label{fig:sim-diasca}
\end{figure}

Figure~\ref{fig:sim-diasca} shows the architecture of the simulator. \ac{Sim-Diasca} (lower layer) is in charge of synchronizing the time between the actors, evolving the system state, sending and receiving messages to and from the controller (i.e., decision-making agent), and managing the results. Its built-in support for distributed simulation enables deploying simulation scenarios over multiple computers. Through the base actor model, own-defined models can be created. Therefore, \textit{DynamicSim} defines an actor model for the replicas, the server, and the load balancer. The traffic generator and the monitor modules act as an interface between the actors in the lower layer and the high-level functions defined in Python through the sub-pub communication schema developed by ZeroMQ. Finally, we design several user-defined simulation scenarios in the topmost layer, including the one presented in Section~\ref{sec:system_model}. 

However, to be able to interact with a \ac{rl} agent, \textit{DynamicSim} should follow the OpenAI Gym~\cite{brockman2016openai} interfaces. Such interfaces communicate the states, actions, and rewards, providing an abstraction layer between the agent and the environment. In this manner, the agent is unaware of how the environment goes from one state to another, and conversely, the environment is unaware of how the agent makes decisions. The interfaces were created using \ac{sb3} custom environments, which inherit the methods from Gym Class that provide that abstraction layer. Thus, in each interaction (or step), the agent chooses an action according to a policy and receives the following state from the simulator. The agent is rewarded or penalized based on the \ac{rfn} depending on the received state. This process is repeated until the agent converges to a policy that maximizes the expected reward in the long term. 

At the level of the simulator, an initial set of actors is generated based on the defined simulation scenario in \ac{Sim-Diasca}. In a simulation scenario, the duration of a step is user-defined. Within a step, the actors simulate its functionality, representing the work done in such a duration. After each actor finishes its simulated work, the system's state is gathered and sent to the agent, the time manager increases the step by one, and the simulation goes to the next step. 

The initial scenario consists of a server with two replicas and a load balancer between them. The load balancer evenly divides the incoming workload among the created replicas. The system state at step $t$ is defined as $s_t = \langle v_t, \bar{c}_t, d_t \rangle$, as explained in Section~\ref{sec:system_model}. Notice that $d_t$ and $\bar{c}_t$ are real variables, while $v_t$ is an integer variable, which poses the scaling problem into the continuous space. The latency is reported since we need to guarantee that each replica can fulfill the \ac{qos} requirement. 

Similarly, at step $ t$, the agent takes action $ A_t = -1, 0, 1$. However, it takes at least one slot to boot or shut down a replica, depending on the algorithm's decision. No job can be assigned to that particular replica during the boot time. Similarly, the replica resources are released during the shutdown time once its pending jobs are processed.

\begin{figure}
    \centering
    \includegraphics[width=0.6\columnwidth]{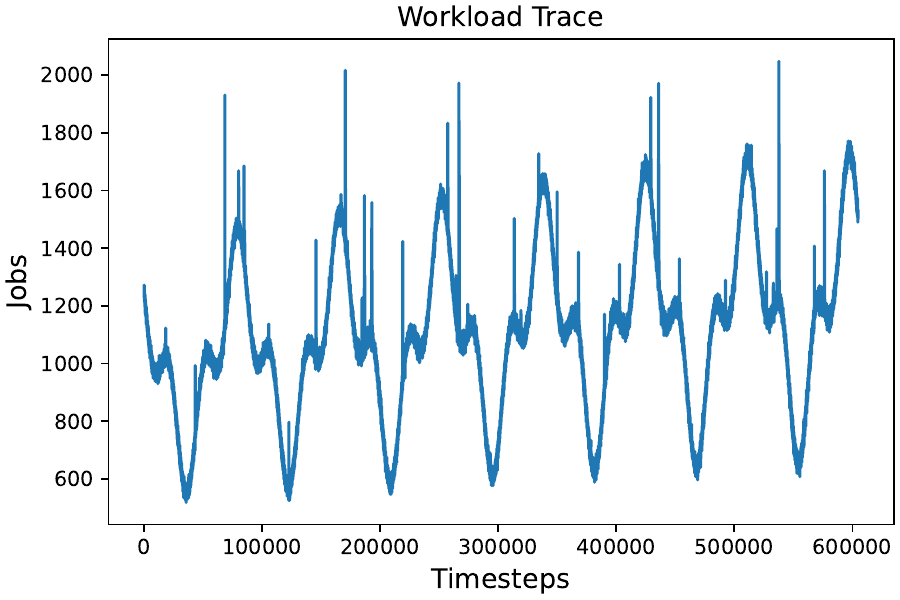}
    \caption{Complete workload trace}
    \label{fig:workload_trace}
\end{figure}

The traffic generator produces a workload following a known pattern in data centers, as shown in Figure~\ref{fig:workload_trace}. Generally, the traffic to a data center is low at night and peaks during working hours. This pattern repeats more or less during the weekday. For more details about how this workload is generated, we refer the reader to~\cite{soto2021towards}. Notice that the workload replicates a dynamic and varying pattern that contains long-term and short-term fluctuations due to sudden changes. The short-term fluctuations are particularly hard to detect in traditional scaling mechanisms since the number of replicas is calculated based on worst-case estimates that do not include such variations. Although synthetic workloads may not fully represent the real incoming traffic, they are suitable for controlled experimentation.

The simulator and the abstraction layer are publicly available\footnote{We are currently working with the legal department at our Lab to make a public version complying with our policy of Open Data. The benchmark will be released before the publication of the manuscript.} since it is within our objectives to promote openness and reproducibility of \ac{rl} algorithms in the networking community, which is fundamental to benchmark \ac{ml} algorithms in networking. 
\section{Training, validating, and selecting DRL algorithms for scaling: A methodology}\label{sec:develop-verify}
In the previous sections, we detailed the influence of scaling computing resources in service-based architectures on system requirements, shaping the learning objectives for autonomous agents (cf. Section~\ref{sec:problem}). Building upon these requirements, Section~\ref{sec:design} delved into the design of the algorithms and the environment, underpinning the realization of autonomic control. This section focuses on the development phase. Assuming the availability of multiple \ac{rl} models for deployment, we incorporate best practices for algorithm comparison and selection, thereby expanding current \ac{mlops} frameworks.

\subsection{Training}\label{sec:develop-verify:training}
Once the appropriate \ac{rl} algorithms and \acp{rfn} tailored to the problem have been selected, the next step in their lifecycle is to develop the autonomous agent. This is commonly known as training. Following the recommendations from the \ac{oran} alliance, \ac{rl} algorithms should be trained offline~\cite{oran-wg2}. Offline training in networking can be performed in a \ac{ndt}~\cite{almasan2022network}, where the outcomes of the untrained agents can be safely injected without compromising the stability of the production network~\cite{camelo2022daemon}. The environment of Section~\ref{sec:design:environment} can serve as a \ac{ndt} for the proposed use case. 

A major challenge in the \ac{rl} community is that minor implementation details can considerably impact performance, sometimes surpassing the disparity between various algorithms~\cite{engstrom2019implementation}. Ensuring the reliability of implementations employed as experimental baselines is paramount. Otherwise, comparing novel algorithms to unreliable baselines may result in inflated estimates of performance enhancements. Thus, standard, open-source frameworks that provide reliable implementations of \ac{rl} algorithms are recommended~\cite{stable-baselines3, achiam2018spinning}. 

Regarding the implementation of the algorithms presented in Section~\ref{sec:design:algorithms}, for this study, we use stable baselines, a popular framework that provides a set of reliable implementations of \ac{rl} algorithms in PyTorch\footnote{\url{https://pytorch.org/}}. We use the default architecture and default hyper-parameters given by stable baselines for both algorithms. Both algorithms use two fully connected networks with 64 units per layer. While actor and critic are shared for \ac{ppo} to reduce computation, \ac{dqn} has separate feature extractors, one for the actor and one for the critic since the best performance is obtained with this configuration. 

Additionally, since \ac{rl}'s reproducibility can be far more difficult than expected~\cite{islam2017reproducibility, henderson2018deep}, given their random initialization of the parameters, among others, Colas et al.~\cite{colas2018many, colas2019hitchhiker} suggest evaluating several seeds of \ac{drl} algorithms in an ofﬂine manner. In this offline evaluation, the algorithm performance after training is independently assessed, usually using a deterministic version of the current policy. We call an \textit{agent} an algorithm - \ac{rfn} - seed combination. 

Consequently, to compare the performance of the different \ac{rl} algorithms, we follow the guidelines suggested by Colas et al.~\cite{colas2018many, colas2019hitchhiker}. In their paper, the authors suggest (i) obtaining the measure of performance of every agent to plot the learning curves of the algorithms, (ii) performing statistical difference testing, (iii) comparing the full learning curves not only comparing the ﬁnal performances of the two algorithms after $t$ steps in the environment. 

\begin{figure}
    \centering
    \includegraphics[width=0.8\textwidth]{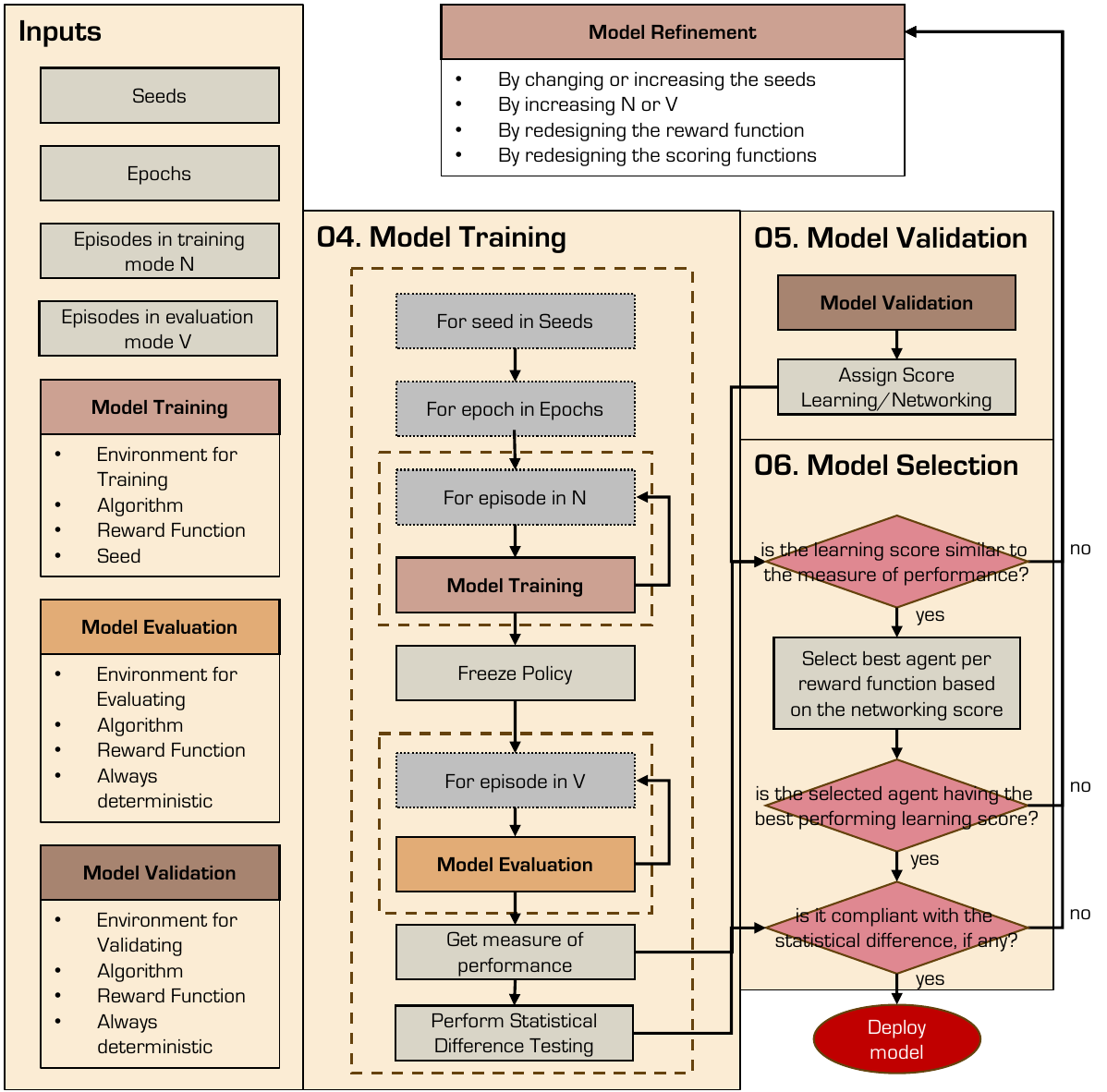}
    \caption{Proposed Methodology}
    \label{fig:methodology}
\end{figure}

Taking those recommendations into account, we extend the current methodological proposals for lifecycle management of \ac{rl} applications in service-based architectures to support algorithmic comparison and selection. We summarize this extension into the methodology shown in Figure~\ref{fig:methodology}. 

At least five different agents should be implemented initially. Each agent is trained using episodes, where an episode is defined as the number of steps until the simulation must be restarted or a maximum number of steps is reached. Accordingly, we determined two situations where an episode is terminated: when the agent creates more replicas than needed or lets the latency go above an inadmissible threshold. These two situations represent an undesired agent behavior and must be penalized. In such situations, the episode ends, the agent receives \num{-100} of reward, and the simulation is restarted. After the simulation is restarted, the initial scenario is again deployed. 

Assuming that what happens in an actual second in every step is simulated, we chose one-hour episodes. Thus, the maximum number of steps is \num{3600}. Notice that the episode length may vary during training since, at the beginning of the training, the agent is expected to take random actions, which can easily lead to episode termination due to undesired behavior. After the agent is trained during $N$ episodes, we freeze its policy and evaluate it during $V$ episodes. Following the logic of \ac{sl} approaches, we split the workload trace into two, one for training and another for evaluation, aiming to test the generalization capabilities of the algorithms to unseen data. The training trace uses the workload's first \num{432e3} values (i.e., first 5 days), while the evaluation trace uses the last \num{172.8e3} values (i.e., the last 2 days). Training and evaluating the agent is considered an epoch; we repeat the process during $E$ epochs. 

\begin{table}[width=.5\linewidth,cols=2,pos=ht]
    \caption{Parameters used in the experimental evaluation.}
    \label{tab:parameters}
    \begin{tabular*}{\tblwidth}{@{} LL@{} }
    \toprule
    \textbf{Parameter}              & \textbf{Value} \\
    \midrule
    Steps per episode               & 3600           \\
    Episodes in training mode $N$   & 24             \\ 
    Episodes in evaluation mode $V$ & 12             \\ 
    Epochs $E$                      & 10             \\ 
    $d_{tgt}$                       & 20 ms          \\ 
    $c_{tgt}$                       & 75\%           \\ 
    $\epsilon$                      & 20\%           \\
    $w_v, w_d$                      & 0.5          \\
    \bottomrule
    \end{tabular*}
\end{table}

Generally, the values of $N$, $V$, and $E$ should be selected by trading off training time and the agent's performance. The agent's performance should be evaluated as soon as possible, which implies selecting a lower training time, i.e., lower $N$, but ensuring enough training is reflected in improving the agent's behavior. Moreover, a larger $V$ gives a better estimation of the true agent's performance but, at the same time, slows down the training. Following this trade-off, we selected the values for $N$, $V$, and $E$ for all \acp{rfn} and \ac{rl} algorithm combinations. The values are shown in Table~\ref{tab:parameters}.

At the end of an evaluation episode, we record the accumulated reward of the episode. Then, the performance per epoch is the average earned reward over the $V$ episodes. To provide a fair comparison of the performance of the algorithms among the different \acp{rfn}, all the \acp{rfn} must be in the same range. Unfortunately, that is not true for the proposed \acp{rfn}. However, a minimum and a maximum achievable reward can be calculated for every agent, depending on the episode duration. 

Suppose an agent makes good decisions to avoid falling into episode termination cases. In that case, the maximum reward possible is the duration of a full episode times the maximum reward. A similar analysis can be done with the minimum achievable reward. Table~\ref{tab:min-max-reward} shows the range variation in each \ac{rfn}, where the underscore in \ac{rfn}3 indicates the optimization profile, c.f., Table~\ref{tab:opt-profiles-rf3}. Having the minimum and maximum range of the reward, we apply a min-max scaler so the agents' performance falls within the same range. 

Once each agent's and run's performance was calculated, we performed statistical difference testing as a main way to compare their performance. In particular, we completed the Welch t-test on the data since it was more robust than other tests to the violation of their assumptions~\cite{colas2019hitchhiker}.

\begin{table}[width=.5\linewidth,cols=3,pos=ht]
    \caption{Minimum and Maximum reward possible in each of the reward functions.}
    \label{tab:min-max-reward}
    \begin{tabular*}{\tblwidth}{@{} CCC@{} }
    \toprule
    \textbf{\begin{tabular}[c]{@{}c@{}}Reward \\ Function\end{tabular}} & \textbf{Min} & \textbf{Max} \\
    \midrule
    RFn1    & 0     & 3600 \\ 
    RFn2    & 0     & 3600 \\ 
    RFn3\_1 & -3600 & 1800 \\ 
    RFn3\_2 & -3600 & 3564 \\ 
    RFn3\_3 & -3600 & 36   \\ 
    \bottomrule
    \end{tabular*}
\end{table}

\subsection{Validation} \label{sec:develop-verify:validation}
Once trained and evaluated, the performance of all the scalers is assessed. This performance can be evaluated from two perspectives: how well the agent learns and how well the agent performs the task at hand. For this purpose, we tested the learned policy in deterministic mode after the $E$ epochs of training by letting the agent run a validation episode of \num{172.8e3} steps. We gathered the accumulated reward the agents received at the end of the validation episode and the main statistical figures of the number of created replicas and the latency. Depending on the obtained values, we score all the agents using Equations~\ref{eq:score-learning} and~\ref{eq:score-networking}. 

\begin{equation}
    \text{score learning} = \frac{\text{normalized accumulated reward}}{\text{normalized episode length}}
    \label{eq:score-learning}
\end{equation}

Equation~\ref{eq:score-learning} scores the agents regarding the learning perspective. The episode termination cases should be avoided if the agent is well-trained. If these cases do not occur, the agent maintains the simulation running, and the episode length is the largest it could be, i.e., \num{172.8e3} steps. Then, using a min-max scaler, the normalized length of a testing episode should be \num{1}, and the score should be approximately the same as the normalized reward achieved during training (see the y-axis of Figure~\ref{fig:results-train}). Suppose the agent falls into the episode termination cases. In that case, the validation episode is finalized before time, where the normalized episode length is lower than one, forcing the score to be higher than one. In this case, the agent can be discarded since it is unacceptable. 

\begin{align} 
\text{score networking} = w_v \cdot V'  + w_d \cdot D' \label{eq:score-networking}\\
V' = 1 - \frac{\bar{v}-min}{max - min}  \label{eq:score-networking:replicas}\\
D' = \frac{\bar{d}}{max} \label{eq:score-networking:latency}
\end{align}

Equation~\ref{eq:score-networking} scores the agents regarding the networking perspective in terms of the number of created replicas and the fulfillment of the \ac{sla} expressed in terms of the processing latency. The goal of a scaler is to fulfill the \ac{sla} with the minimum number of replicas needed. Then, to compare how an agent performs over the remaining agents, we need also to apply normalization. For the number of replicas, we use min-max normalization as expressed in Equation~\ref{eq:score-networking:replicas}, where $\bar{v}$ is the average number of replicas created during the validation episode, $min$, and $max$ are the minimum and the maximum values of all the created agents. In this case, we applied an inverse normalization since the min-max normalization will return values close to zero when $\bar{v}$ is close to the minimum. 

Similarly, Equation~\ref{eq:score-networking:latency} normalizes the latency. By using only the maximum latency value, Equation~\ref{eq:score-networking:latency} is able to identify how far this maximum is from its mean value. The closer to one, the distribution is more centered around the mean, while the closer to zero, the distribution presents a right-hand side tail. This type of normalization is handy for detecting extreme peak values in the latency distribution~\cite{rochet2016mean}, which should be avoided in mission-critical applications, for instance. Notice that, while the $min$, $max$ in the previous equation should select the minimum (or maximum) between all the created agents, the $max$ in Equation~\ref{eq:score-networking:latency} corresponds to the maximum latency value of each agent during the validation episode. This value is associated with the latency distribution of the agent itself, and therefore it is irrelevant to compare it against other agents. 

Being $V'$ the normalized value of the replicas created by a given agent and $D'$ the normalized latency of a given agent, Equation~\ref{eq:score-networking} tries to find a balance between the two main objectives of the scaler. $w_v$ and $w_d$ are weights used to tune which of the two criteria is more important to the \ac{nsp}. Moreover, the equations are chosen depending on the service type that the \ac{nf} is providing. For instance, if the application allows the users to experiment some extreme latency issues, then Equation~\ref{eq:score-networking:latency} could be replaced by another equation, e.g., $\bar{d} / d_{tgt}$ measures how far are the mean latency values from the target. After all the agents have been scored, model selection can start. 

\subsection{Selection}\label{sec:develop-verify:selection}
Model selection is a filtering process in which all the agents that were not able to perform in a plausible way during the previous two stages are ruled out. The first step in this process is to discard all the agents whose learning score is not similar to the measure of performance obtained during the training. As explained above, a well-trained agent is able to maintain the running of the validation episode; if this is not the case, the learning score will be higher than the measure of performance, indicating a possible outlier. Such agents should undergo model refinement, by, e.g., changing the seed. We filter first by the learning score since an ill-trained agent is most likely to have a poor performance in the scaling task. 

Then, by using the networking score, we select the best scoring agent per reward function. Ultimately, the agent's true performance is given in terms of the networking \acp{kpi} and not the reward the agent achieves during training. The chosen agent must satisfy two criteria: firstly, its learning score should rank among the highest within its respective \ac{rfn}; secondly, if there exists statistical evidence demonstrating higher performance of one algorithm over another, the highest-performing agent should be trained using the algorithm proven to be more effective. If at least one of the criteria is met, the agent with the highest learning score should be deployed. In case none of the criteria is met, the agents should undergo model refinement, by either increasing the number of agent-environment interactions or by redesigning the \ac{rfn}.
\section{Experimental evaluation}\label{sec:results}
\begin{figure*}
    \centering
    \subfloat[Learning curve of \ac{rfn}1\label{fig:train-rf1}]{\includegraphics[width=0.3\textwidth]{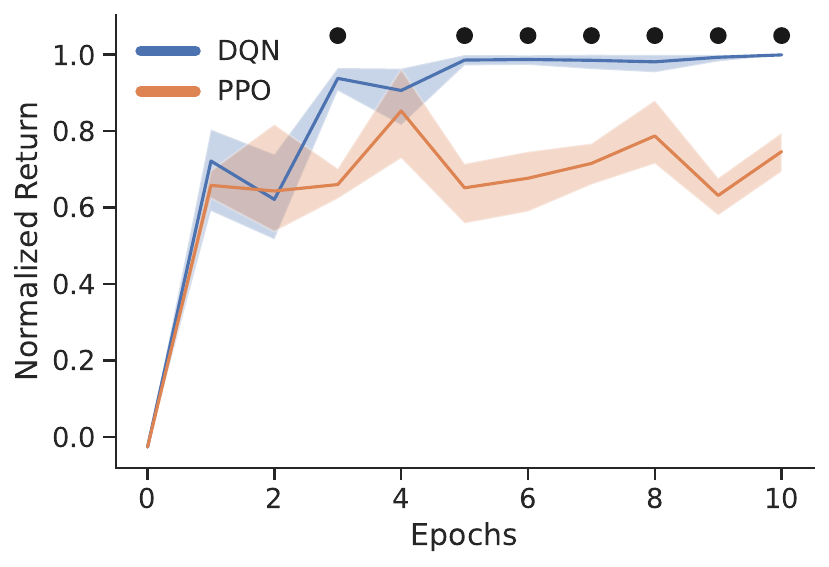}}
    \subfloat[Learning curve of \ac{rfn}2\label{fig:train-rf2}]{\includegraphics[width=0.3\textwidth]{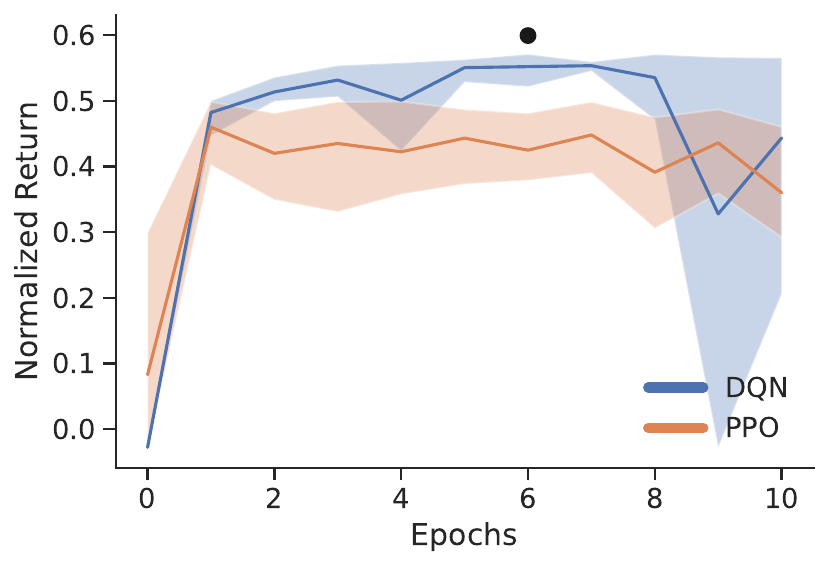}}
    \\
    \subfloat[Learning curve of \ac{rfn}3\_1\label{fig:train-rf3-1}]{\includegraphics[width=0.3\textwidth]{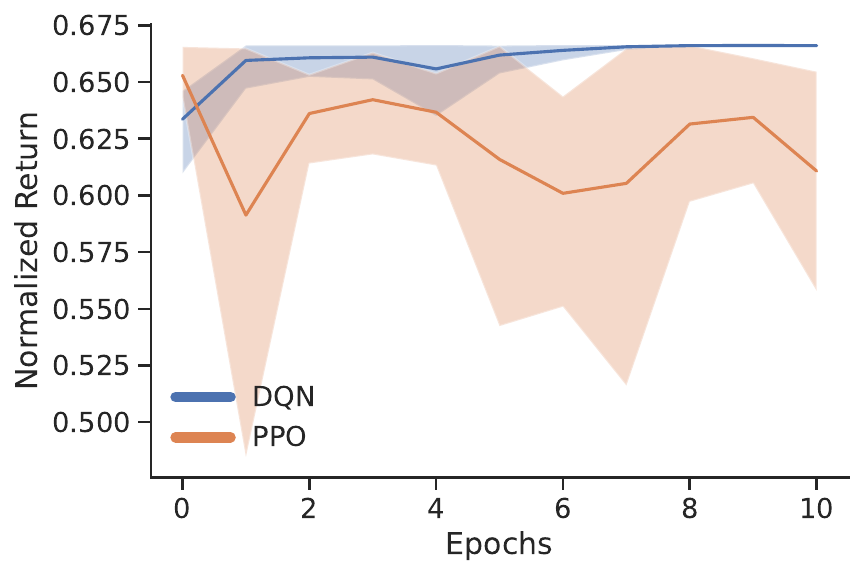}}
    \subfloat[Learning curve of \ac{rfn}3\_2\label{fig:train-rf3-2}]{\includegraphics[width=0.3\textwidth]{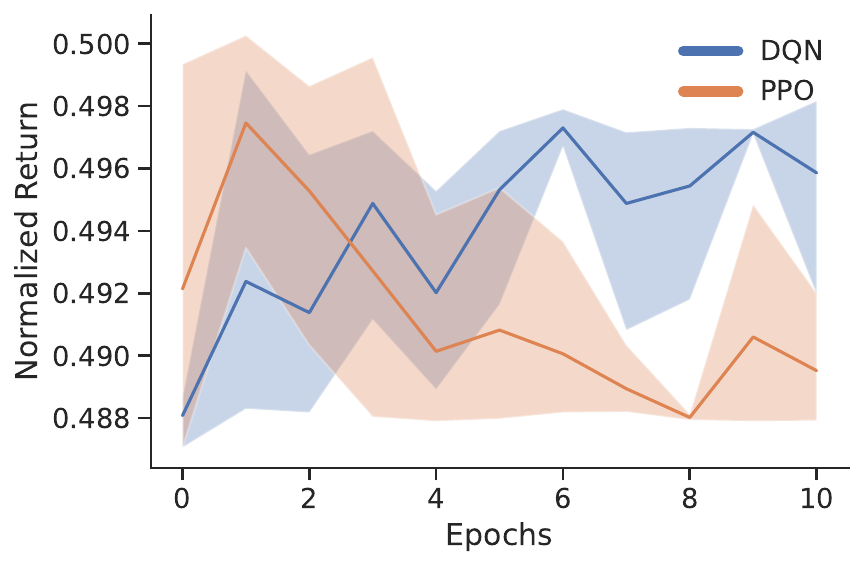}}
    \subfloat[Learning curve of \ac{rfn}3\_3\label{fig:train-rf3-3}]{\includegraphics[width=0.3\textwidth]{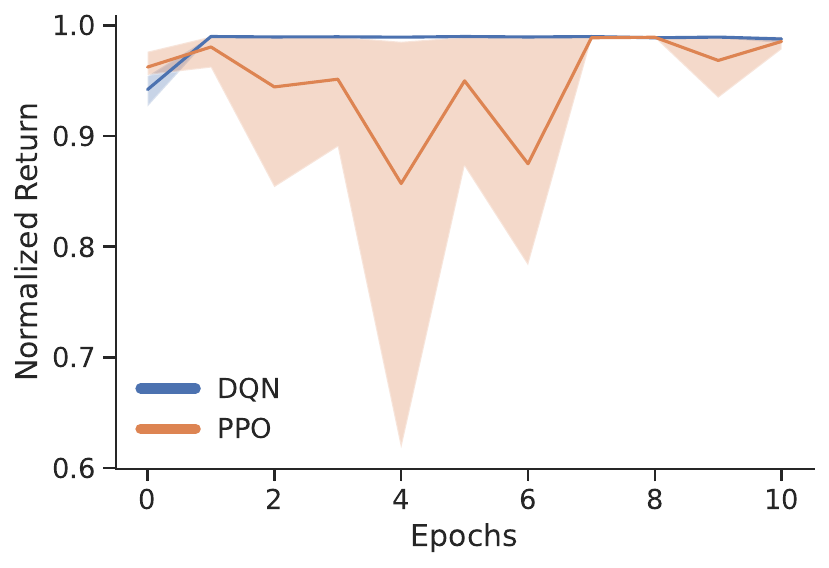}}
    \caption{Learning curves, with error shades and black dots indicating significant statistical difference when present.}
    \label{fig:results-train}
\end{figure*}

This section presents the experimental evaluation of the two \ac{drl} algorithms and the associated \acp{rfn}. Through a systematic assessment, we aim to elucidate the impact of these algorithmic and \ac{rfn} choices on the learning dynamics, convergence speed, and overall effectiveness of our proposed \ac{rl} methodologies. The experiments are designed to provide insightful comparisons, shedding light on the strengths and limitations of each algorithm and offering a comprehensive understanding of their behavior. This empirical analysis validates our contributions and adds to the broader discourse on designing and optimizing \ac{rl} algorithms for real-world applications.

\subsection{Training performance}
Using the procedure described in Section~\ref{sec:develop-verify:training}, we trained and validated the algorithms of Section~\ref{sec:design:algorithms} using the \acp{rfn} described in Section~\ref{sec:design:rfs}. Figure~\ref{fig:results-train} shows the learning curves, with 80\% error percentile and black dots indicating statistical significance when applicable for all \acp{rfn}. The agents were also evaluated before training to see how much improvement that learning brings. This is plotted as epoch \num{0}. 

As it can be observed from the figure, only \ac{rfn}1 shows consistent statistical difference across the epochs, where the \ac{dqn} performs better than the \ac{ppo}. Unfortunately, we cannot draw any conclusions for the remaining \acp{rfn}; this may be due to the requirement for additional data. More data can be gathered using two different methods. On the one hand, having more seeds will enable averaging more trials, yielding a more reliable algorithm performance metric~\cite{colas2018many}. On the other hand, for some cases, e.g., Figures~\ref{fig:train-rf3-1},~\ref{fig:train-rf3-2},~\ref{fig:train-rf3-3}, it is noticeable that the algorithms, especially the \ac{ppo}, have not yet converged to a stable value of the reward. More training epochs can help the algorithm reach convergence in these cases. 

Notice, however, that when we talk about convergence, we do not talk about the algorithm converging to an optimal policy but to a stable reward value, which a sub-optimal policy can produce. Regarding the convergence to an optimal policy, it is known that \ac{drl} approaches are sample inefficient~\cite{yu2018towards} and that, for instance, Q-learning algorithms are guaranteed to converge in the limit, when each state-action pair is visited infinitely often~\cite{watkins1992q}. Some strategies have been incorporated in several state-of-the-art algorithms, such as using a $\varepsilon$-greedy policy where $\varepsilon > 0$ or by using a learning rate close to zero. However, in practice, \ac{drl} algorithms require millions of samples to achieve good and stable performance depending on the task~\cite{schulman2017proximal, mnih2015human}. 

In our problem, we trained and evaluated each algorithm during $N$ and $V$ episodes (see Table~\ref{tab:parameters}), where each episode consists, in the best case, of $3600$ samples or interactions. Then, we have $3600 \times V \times E =$ \num{432e3} samples or $3600 \times (N + V) \times E =$ \num{1296e3} samples, if we consider the training interactions. The number of samples we used is lower than those conventionally used in \ac{rl}. Nevertheless, both \ac{drl} algorithms are able to stabilize when using \ac{rfn}1 (Figure~\ref{fig:train-rf1}), and to some extent, using \ac{rfn}2 (Figure~\ref{fig:train-rf2}), while the \ac{dqn} is also able to stabilize using \ac{rfn}3\_1 (Figure~\ref{fig:train-rf3-1}) and \ac{rfn}3\_3 (Figure~\ref{fig:train-rf3-3}). 

\begin{table}[width=.9\linewidth,cols=11,pos=ht]
\caption{Learning score of all the individual runs. In red are highlighted the cases where the agents terminated the validation episode before time.}
\label{tab:score-learning}
\begin{tabular*}{\tblwidth}{@{} LCCCCC|CCCCC@{} }
\toprule
    & \multicolumn{5}{C|}{\textbf{DQN}} & \multicolumn{5}{C}{\textbf{PPO}} \\
    & \textbf{Run 1} & \textbf{Run 2} & \textbf{Run 3} & \textbf{Run 4} & \textbf{Run 5} & \textbf{Run 1} & \textbf{Run 2} & \textbf{Run 3} & \textbf{Run 4} & \textbf{Run 5} \\
\midrule
\textbf{RFn1} & 0.9971 & 0.9999 & 0.9980 & 0.9999 & 0.9999 & 0.7462 & 0.7427 & 0.6765 & 0.6976 & 0.7529 \\ 
\textbf{RFn2} & 0.5662 & 0.5565 & 0.5521 & 0.5629 & 0.5598 & 0.3777 & 0.3784 & 0.3781 & 0.3784 & 0.3852 \\
\textbf{RFn3\_1} & 0.6666 & 0.6666 & 0.6666 & 0.6666 & 0.6666 & \color[HTML]{FE0000}246.7 & 0.6581 & \color[HTML]{FE0000}1.522 & 0.6567 & 0.6588 \\
\textbf{RFn3\_2} & 0.4975 & \color[HTML]{FE0000}22.28 & 0.4975 & 0.4975 & 0.4975 & \color[HTML]{FE0000}1.181 & \color[HTML]{FE0000}768.0 & \color[HTML]{FE0000}340.3 & \color[HTML]{FE0000}594.4 & \color[HTML]{FE0000}22.18 \\
\textbf{RFn3\_3} & \color[HTML]{FE0000}43.92 & 0.9901 & 0.9901 & 0.9901 & 0.9901 & 0.9901  & 0.9901 & 0.984 & 0.9894 & 0.989 \\ 
\bottomrule
\end{tabular*}
\end{table}

The value around each agent stabilizes differs from \ac{rfn} to \ac{rfn}. This is because of how often they achieve a reward in their trajectory. In \ac{rl}, a trajectory refers to a sequence of states, actions, and rewards an agent experiences while interacting with an environment over a certain period. Trajectories are fundamental to \ac{rl} as they are used to learn and update the agent's policy or value function, enabling it to make better decisions based on the cumulative experience gained during these trajectories. In the following Section, we will look closely at each \ac{rfn}, and via the results from the testing phase, we will analyze how each \ac{rfn} influences the agent's behavior in the number of created replicas and the latency control.

\begin{table}[width=.9\linewidth,cols=11,pos=ht]
\caption{Networking score of all the individual runs. The red dash identifies the agents disregarded in the previous stage, while in green are shown the best-performing agents per RFn.}
\label{tab:score-networking}
\begin{tabular*}{\tblwidth}{@{} LCCCCC|CCCCC@{} }
\toprule
    & \multicolumn{5}{C|}{\textbf{DQN}} & \multicolumn{5}{C}{\textbf{PPO}} \\
    & \textbf{Run 1} & \textbf{Run 2} & \textbf{Run 3} & \textbf{Run 4} & \textbf{Run 5} & \textbf{Run 1} & \textbf{Run 2} & \textbf{Run 3}          & \textbf{Run 4} & \textbf{Run 5} \\
\midrule
\textbf{RFn1} & 0.4142 & 0.4467 & 0.4166 & \color[HTML]{009901} 0.4591 & 0.4567 & 0.4525 & 0.4519 & 0.4496 & 0.4444 & 0.4513 \\
\textbf{RFn2} & 0.5219 & 0.5225 & \color[HTML]{009901} 0.5265 & 0.4472 & 0.4455 & 0.4867 & 0.4968 & 0.4883 & 0.5005 & 0.4931 \\
\textbf{RFn3\_1} & 0.4499 & \color[HTML]{009901} 0.5062 & 0.4999 & 0.4774 & 0.4736 & \color[HTML]{FE0000} - & 0.3886 & \color[HTML]{FE0000} - & 0.4193 & 0.3894 \\ 
\textbf{RFn3\_2} & 0.8695 & \color[HTML]{FE0000} - & 0.7842 & 0.8863 & \color[HTML]{009901} 0.8873 & \color[HTML]{FE0000} - & \color[HTML]{FE0000} - & \color[HTML]{FE0000} - & \color[HTML]{FE0000} - & \color[HTML]{FE0000} - \\
\textbf{RFn3\_3} & \color[HTML]{FE0000} - & 0.3026 & 0.2577 & 0.3462 & 0.3977 & 0.4416 & \color[HTML]{009901} 0.4630 & 0.4190 & 0.4289 & 0.3844 \\ 
\bottomrule
\end{tabular*}
\end{table}

\subsection{Validation performance}
As mentioned in Section~\ref{sec:develop-verify:validation}, we run each agent in a validation episode. The scores of each run are shown in Tables~\ref{tab:score-learning}, and~\ref{tab:score-networking}. Regarding the learning score, we observe in Table~\ref{tab:score-learning}, the cases where this score is above \num{1}, highlighted in red. Such cases, are immediately discarded and are not scored using Equation~\ref{eq:score-networking}. According to the methodology introduced in Section~\ref{sec:develop-verify} (see Figure~\ref{fig:methodology}), such agents can be replaced by others by changing the seed, for instance. Besides that fact, for most cases, the learning score corresponds to the stabilizing value of the normalized reward in their learning curves, as shown in Figure~\ref{fig:results-train}. In general, the learning score of the agents per \ac{rfn} and algorithm is similar, which leads us to think that the algorithms can find an akin policy independently of their weight initialization.  

Regarding the networking performance, as depicted in Table~\ref{tab:score-networking}, agents that were eliminated in the previous stage are denoted with a red dash, while those with the highest scores per \ac{rfn} are highlighted in green. Similar to Table~\ref{tab:score-learning}, scoring is consistent among agents trained with the same \ac{rfn}. However, our two-step approach, as outlined in our methodology, enables the identification of the most suitable model for deployment. 

For instance, suppose we adhere to the conventional method of selecting the agent with the highest accumulated reward. In that scenario, agents \num{2}, \num{4}, and \num{5} of the \ac{dqn} utilizing \ac{rfn}1 could all be considered viable options. Nevertheless, upon closer examination of their scores, it becomes apparent that agent \num{2} slightly underperforms compared to the others. Furthermore, as illustrated in Table~\ref{tab:score-networking}, \ac{ppo} agents demonstrate better-than-anticipated performance, contrary to expectations based solely on their learning scores.

Combining both tables reveals that, in certain instances, assumptions based on learning scores align with evidence from networking scores. For example, agents trained with \ac{rfn}1 consistently outperform \ac{ppo} agents in both learning and networking. Similar trends are observed in \ac{rfn}2. In the case of \ac{rfn}3\_1, although both algorithms exhibit similar learning performance, \ac{dqn} demonstrates clear superiority over \ac{ppo} in networking. Conversely, in \ac{rfn}3\_3, while all agents exhibit similar learning scores, networking scores indicate better performance by \ac{ppo} agents for this specific function. In the following section, we look more closely at the behavior of the agents per \ac{rfn}, including the best-performing ones. 

\subsection{Selection}

In the final selection phase, we identify the top-performing agents as follows: (i) \ac{dqn} \textit{Run \num{4}} for \ac{rfn}1, (ii) \ac{dqn} \textit{Run \num{3}} for \ac{rfn}2, (iii) \ac{dqn} \textit{Run \num{2}} for \ac{rfn}3\_1, (iv) \ac{dqn} \textit{Run \num{5}} for \ac{rfn}3\_2, and (v) \ac{ppo} \textit{Run \num{2}} for \ac{rfn}3\_3. According to our methodology, we next verify if these agents meet the deployment criteria.

The learning score criterion ensures the selected agents are among the highest-ranked within their respective \acp{rfn}. As observed from Tables~\ref{tab:score-learning} and~\ref{tab:score-networking}, only \ac{dqn} \textit{Run \num{4}} with \ac{rfn}1 and \ac{ppo} \textit{Run \num{2}} with \ac{rfn}3\_3 achieve high scores in both learning and networking. Further analysis of the learning curves in Figure~\ref{fig:results-train} consistently shows that \ac{dqn} agents trained with \ac{rfn}1 outperform \ac{ppo} agents, validating the selection of \ac{dqn} \textit{Run \num{4}} for deployment.

Table~\ref{tab:best-agents} supports the observation that \ac{dqn} generally outperforms \ac{ppo}. The \ac{ppo} agent tends to stabilize around \num{7} replicas, maintaining compliance with the \ac{sla} with minimal violations. In contrast, the \ac{dqn} agent displays more variability in replica creation with slightly increased \ac{sla} violations than the \ac{ppo} counterpart. While the \ac{ppo} agent might be preferred in scenarios requiring consistent replica creation, the \ac{dqn} agent is more effective in balancing deployment costs with user \ac{qos}. This makes \ac{dqn} generally more suitable than \ac{ppo} for this task.

\begin{table}[width=.7\linewidth,cols=5,pos=h]
\caption{Main Statistical values for the best-performing agents}
\label{tab:best-agents}
\begin{tabular*}{\tblwidth}{@{} LCC|CC@{} }
\toprule
    & \multicolumn{2}{C|}{\textbf{Replicas}} & \multicolumn{2}{C}{\textbf{Latency}} \\ 
    & \textbf{RFn1} & \textbf{RFn3\_3} & \textbf{RFn1} & \textbf{RFn3\_3} \\
    & \textbf{DQN - Run 4} & \textbf{PPO - Run 2} & \textbf{DQN - Run 4} & \textbf{PPO - Run 2}\\ 
    \midrule
\textbf{mean} & \textbf{5.0771} & \textbf{6.9999} & 0.0089          & 0.0077            \\
\textbf{std}  & 1.1276          & 0.0215          & 0.0004          & 0.0008            \\ 
\textbf{min}  & 2.0000          & 2.0000          & 0.0058          & 0.0058            \\
\textbf{25\%} & 5.0000          & 7.0000          & 0.0087          & 0.0074            \\ 
\textbf{50\%} & 5.0000          & 7.0000          & 0.0089          & 0.0077            \\
\textbf{75\%} & 6.0000          & 7.0000          & 0.0090          & 0.0081            \\
\textbf{max}  & \textbf{9.0000} & \textbf{7.0000} & \textbf{0.0612} & \textbf{0.0297}   \\ 
\bottomrule
\end{tabular*}
\end{table}

\section{Discussion}\label{sec:discussion}
Future \ac{zsm} approaches will require an intelligence orchestrator~\cite{chatzieleftheriou2023orchestration} that autonomously (i) determines when a \ac{ml} algorithm is ready for deployment, (ii) between two or more algorithms, compares which algorithm performs better for a given task, and specifically for \ac{rl} algorithms, (iii) assess the performance of two or more \acp{rfn}. For doing that, a common approach is to compare and benchmark \ac{drl} algorithms based on their reward in a well-defined and standard environment. However, this section will present strong evidence that ranking an algorithm based solely on the return might be insufficient. Therefore, the methodology presented in Section~\ref{sec:develop-verify} tries to provide an automatic way of determining if two agents perform the same or differently or to select a better-performing algorithm, facilitating the work of the intelligence orchestrator. We analyze the agents' behavior per \ac{rfn} to dive into the discussion.

\subsection{Analysis of the behavior of the agents using RFn1}

\begin{table}[width=.5\linewidth,cols=5,pos=ht]
\caption{Main Statistical values of Run 1 of the DQN and the PPO using RFn1.}
\label{tab:results_RF1}
\begin{tabular*}{\tblwidth}{@{} CCCCC@{} }
\toprule
    & \multicolumn{2}{c}{\textbf{Replicas}} & \multicolumn{2}{c}{\textbf{Latency}} \\
    & \textbf{DQN}    & \textbf{PPO}    & \textbf{DQN}    & \textbf{PPO}   \\
    & \textbf{Run 1}  & \textbf{Run 1}  & \textbf{Run 1}    & \textbf{Run 1} \\
\midrule
\textbf{mean}   & \textbf{5.1889}   & \textbf{5.1234}   & 0.0088    & 0.0088  \\
\textbf{std}    & 1.2358            & 0.5913            & 0.0011    & 0.0012  \\
\textbf{min}    & 1.0000            & 2.0000            & 0.0058    & 0.0058  \\
\textbf{25\%}   & 4.0000            & 5.0000            & 0.0086    & 0.0085  \\
\textbf{50\%}   & 5.0000            & 5.0000            & 0.0087    & 0.0089  \\
\textbf{75\%}   & 6.0000            & 5.0000            & 0.0090    & 0.0091  \\
\textbf{max}    & \textbf{9.0000}   & \textbf{6.0000}   & \textbf{0.1440}   & \textbf{0.0657} \\
\bottomrule
\end{tabular*}
\end{table}

This function rewards the agent if the monitored latency $d_t$ or \ac{cpu} usage $c_t$ are between predetermined boundaries. These boundaries are set by defining an operating target and a tolerance range. For this \ac{rfn}, we assume a target latency of \num{20}\si{\milli\second} and a \ac{cpu} usage target of \num{75}\% with  a tolerance of \num{20}\%, as indicated in Table~\ref{tab:parameters}. Notice that the target latency can be established by the \ac{sla} while the \ac{cpu} target needs to be calculated according to operational needs. Figure~\ref{fig:train-rf1} shows that only \ac{rfn}1 provides statistical evidence to support the fact that \ac{dqn} agents are behaving better than the \ac{ppo} ones.

If only the reward is taken into account, Table~\ref{tab:score-learning} suggests that all \ac{dqn} agents should outperform the \ac{ppo} ones. However, Table~\ref{tab:results_RF1} proves this is not the case. The networking score shown in Table~\ref{tab:score-networking} is more accurate in estimating the performance of the agents, where \ac{ppo} \textit{Run 1} performs even better than \ac{dqn} \textit{Run 1} by creating fewer replicas in average and having similar latency control with fewer \ac{sla} violations. 

\begin{figure}
    \centering
    \subfloat[Run 4 of DQN\label{fig:reward-rf1-dqn}]{\includegraphics[width=0.45\columnwidth]{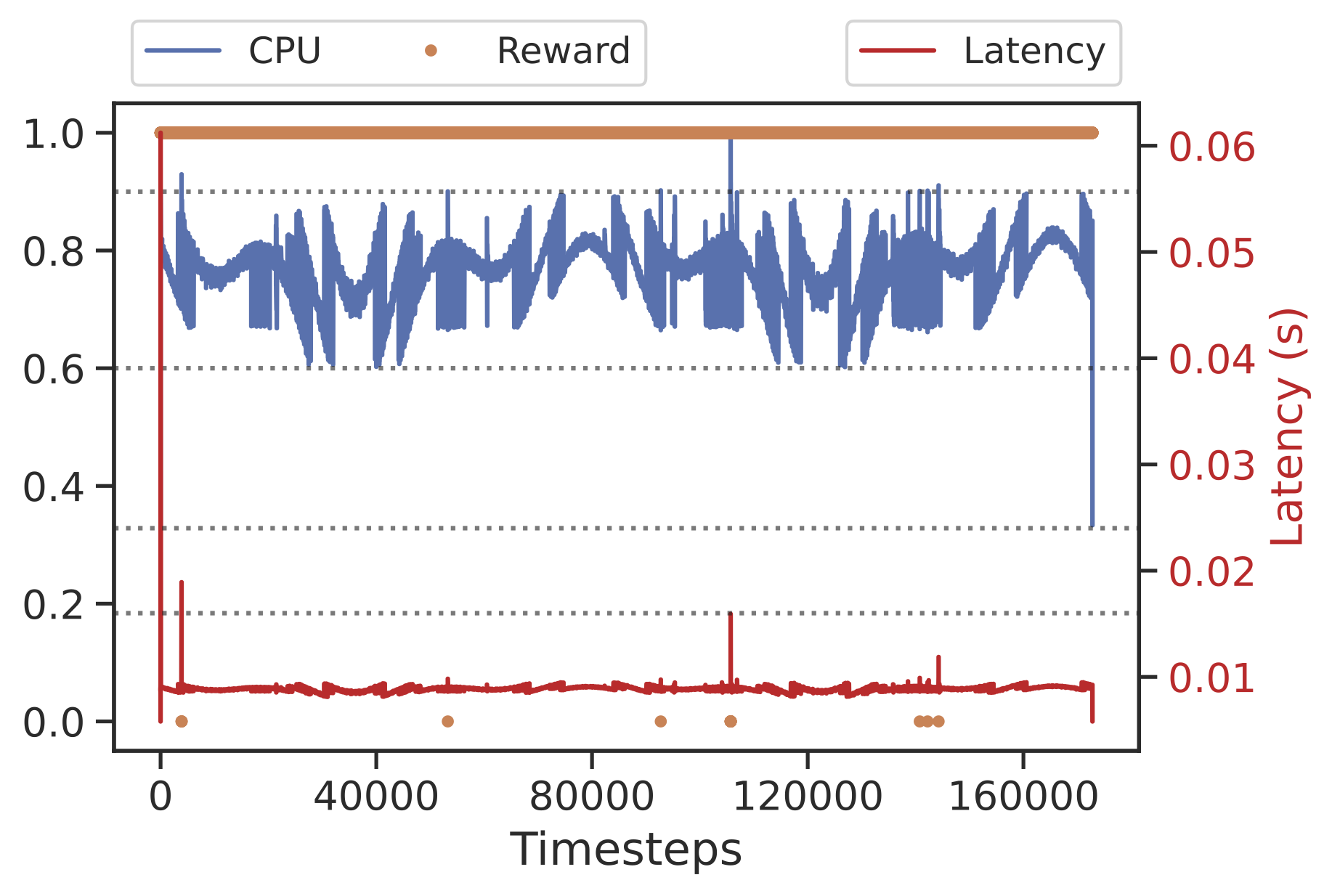}}
    \hfill
    \subfloat[Run 4 of PPO\label{fig:reward-rf1-ppo}]{\includegraphics[width=0.45\columnwidth]{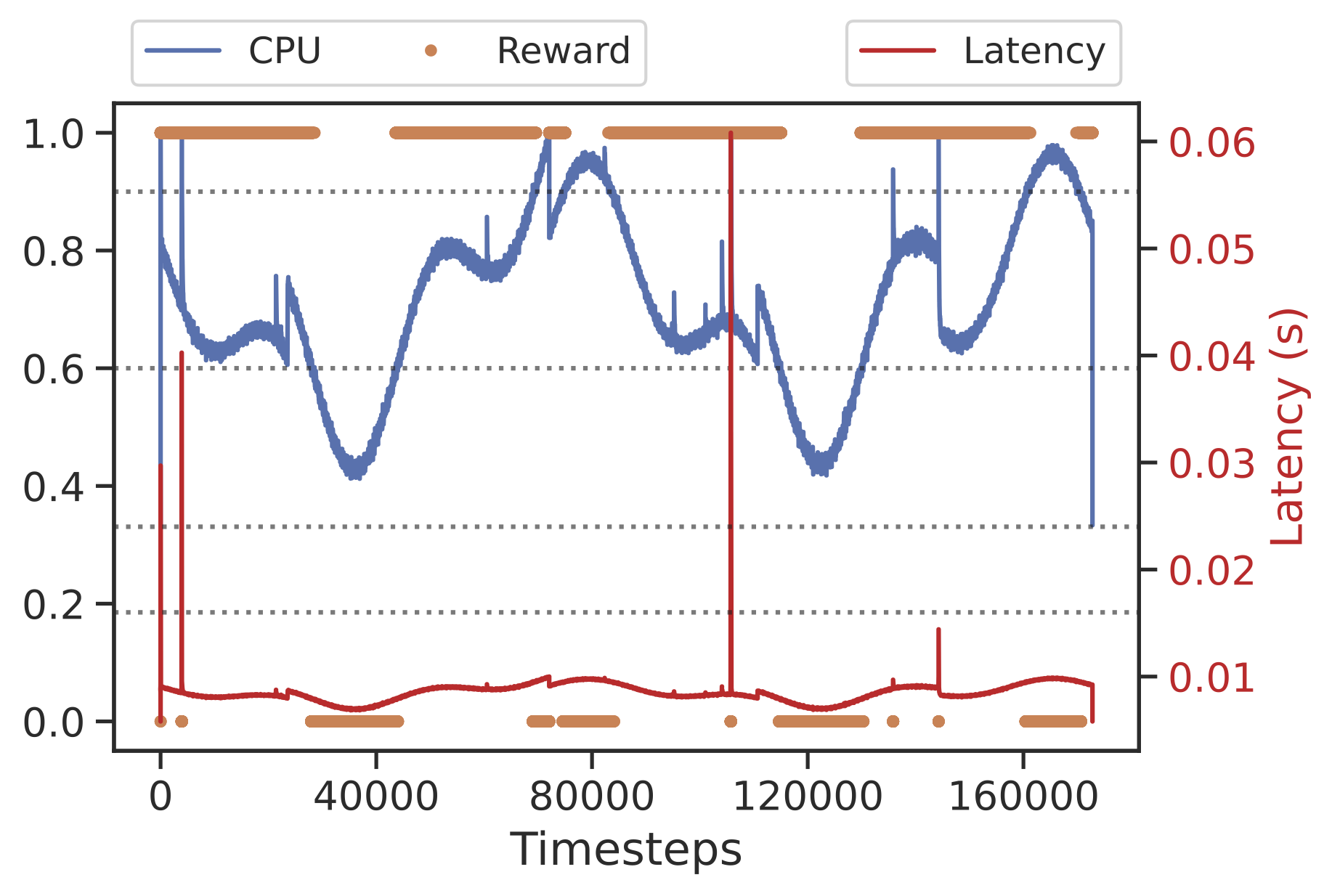}}
    \caption{Different agent behavior using \ac{rfn}1. In blue is the \ac{cpu} usage; in orange is the achieved reward per step; in red is the latency. The black-dotted line shows the respective thresholds}
    \label{fig:reward-rf1}
\end{figure}

Despite their lower learning scores, the \ac{ppo} agents demonstrate performance comparable to \ac{dqn} agents with higher learning scores. The learning score, based solely on the obtained return, indicates that \ac{dqn} algorithms are better at optimizing the objective for this \ac{rfn}, i.e., keep the latency and \ac{cpu} usage within bounds. In contrast, the networking score shows how the agents will perform when deployed. For instance, Figure~\ref{fig:reward-rf1} showcases the top-performing \ac{dqn} agent (\textit{Run 4}) and \ac{ppo} agent (\textit{Run 4}), which has one of the lowest learning scores for \ac{rfn}1. In the graph, the orange line represents the reward achieved by the agent at each step, while the blue and red lines denote the average \ac{cpu} usage and peak latency of created replicas, respectively. Horizontal black dotted lines indicate \ac{cpu} and latency bounds. Notably, the \ac{dqn} agent consistently receives rewards at each step, while the \ac{ppo} agent struggles, particularly during low or high workload periods. As shown in the Figure, the success of \ac{dqn} agents lies in their superior control of \ac{cpu} usage. 

\subsection{Analysis of the behavior of the agents using RFn2}
We assume the same latency target for this \ac{rfn} as for \ac{rfn}1, \num{20}\si{\milli\second}. Consequently, the \textit{Above} state is defined as $d_t$ above the latency target, while the \textit{Below} state is defined as $d_t$ below that target. This \ac{rfn} showcases interesting characteristics of \ac{drl} algorithms. First, this \ac{rfn} is evidence of what in the \ac{rl} community is known as reward hacking~\cite{amodei2016concrete}. Reward hacking is the term employed when the agent finds a way to maximize the expected return that is not aligned with the desired behavior. In particular, some of the trajectories of the agents using \ac{rfn}2 reflect that once an agent is below the latency, it reduces the number of replicas, so it slightly goes above the latency. Being there, the agent increases the number of replicas to receive the reward. This behavior is repeated several times during a trajectory, allowing the agent to receive a reward every 2 or 3 steps with the side effect of occasionally trespassing the latency target. 

Examining the agent behavior more closely, the \ac{ppo} agents adjust replica numbers to transition between states, initially increasing replicas to reach the \textit{Below} state, then reducing them to obtain rewards, before transitioning back to the \textit{Above} state. This cyclic process leads to a gradual increase in achieved rewards over time. Conversely, \ac{dqn} agents employ a similar strategy but exploit the absence of penalties for creating surplus replicas, remaining in the \textit{Below} state with excess replicas to increase the likelihood of reward acquisition. However, both agent types struggle to meet \ac{sla} requirements, with approximately more than \num{20}\% of the time spent in violation, indicating the need for a more refined \ac{rfn} design.

Second, since the agents cannot be rewarded in every step, only a certain combination of actions will produce a reward, which can be considered a sparse \ac{rfn}. However, finding a suitable policy within a reasonable time is more difficult in a sparse reward setting than in a dense one~\cite{eschmann2021reward}. The \ac{rfn}'s sparsity is also why the learning score of the \ac{dqn} and \ac{ppo} agents is lower than their counterparts using \ac{rfn}1. Finally, Figure~\ref{fig:train-rf2}, and Tables~\ref{tab:score-learning} and~\ref{tab:score-networking}, suggest that \ac{dqn} agents perform subtly better than the \ac{ppo} ones. Still, this is not the case for every epoch, especially towards the final ones where the \ac{dqn} agents seem to suffer catastrophic forgetting. 

Notice that, from the perspective of the methodology, all \ac{ppo} agents should be discarded since their learning score (Table~\ref{tab:score-learning}) is higher than the measure of performance (Figure~\ref{fig:train-rf2}). Continuing with the methodology, the best \ac{dqn} agent (\textit{Run 3}) does not have the best-performing learning score (cf. Table~\ref{tab:score-learning}). Therefore, there is no guarantee that any of the \ac{dqn} agents will perform consistently when deployed in production networks.

\subsection{Analysis of the behavior of the agents using RFn3}
This \ac{rfn} is a multi-objective function where depending on the optimization profiles (cf. Table~\ref{tab:opt-profiles-rf3}), an agent tries to minimize the associated cost of creating replicas or surpassing a threshold in latency. In the first optimization profile, the agents will try to balance the two objectives; in the second, the agents will optimize replicas creation; in the last optimization profile, the agents will pay more attention to latency control. Therefore, three different behaviors are expected. Having multiple optimization profiles, a network solution designer can easily switch the agents' behavior so they optimize one objective, i.e., control the latency, or the other, i.e., limit the creation of replicas. 

This \ac{rfn} emphasizes the need for an alternative scoring metric for \ac{drl}-based scalers. Notice how, according to Table~\ref{tab:score-learning}, all \ac{dqn} agents score the same for the different optimization profiles defined by \ac{rfn}3; the networking score helps in determining which agent is actually performing the best, same as in \ac{rfn}1. Moreover, focused only in \ac{rfn}3\_3, the networking score helps in determining the best-performing agent even across algorithms, revealing that the \ac{ppo} agent of \textit{Run 2} performs the best.

Following the methodology reveals that, in \ac{rfn}3\_1, the \ac{ppo} agents are discarded since their learning score is not similar to the measure of performance (cf. Figure~\ref{fig:train-rf3-1}). Moreover, the agents trained with \ac{rfn}3\_2 are discarded since we cannot guarantee that their agents are performing consistently, similar to \ac{rfn}2. 

\subsection{Final Remarks}
From analyzing in more detail the results obtained, we identified that 
achieving a stable reward during training is recommended for \ac{rl} scaling algorithms but not mandatory for an agent to meet key scaling objectives, such as maintaining \ac{sla} compliance with minimal replicas. To address this challenge, we proposed a methodology that includes a scoring system based on the reward to quickly identify efficient agents, which generally corresponds to the stabilized reward values seen in training. 

However, the scoring system has limitations, particularly in distinguishing between agents that appear similar in learning but differ in deployment suitability. To address this, our methodology also evaluates the performance of scaling tasks, focusing on the number of replicas created and \ac{sla} compliance. Despite the lack of a formal evaluation methodology for auto-scaling techniques, our approach aims to fill this gap, particularly in model selection for autonomous network operations.

Finally, we emphasize the critical role of the \ac{rfn} definition in \ac{drl}-based scaling algorithms. While the default \ac{rl} objective is to maximize expected reward, timely decisions in scaling, such as when to create or delete replicas, can significantly impact long-term performance. This explains why agents that excel in managing replicas and latency may receive lower rewards than those with less effective performance. Using \acp{rfn} with non-cumulative objectives~\cite{cui2023reinforcement} may yield better results, as different optimization profiles can lead to varying behaviors that better balance the trade-offs between objectives.
\section{Related work} \label{sec:sota}
This section explores the foundational literature in two main areas: \ac{rl} techniques for scaling in service-based network architectures and benchmarks in \ac{ml} and \ac{rl} within networking.  This examination of related work establishes a contextual framework, paving the way for the contributions and differentiators of our research.

\subsection{Reinforcement Learning for Scaling}

Scaling strategies in computing involve balancing two conflicting objectives: meeting \ac{qos} and \ac{qoe} demands by allocating more resources, and minimizing \ac{opex} and \ac{capex}. Scaling is key to managing resources efficiently while maintaining service quality. There are two main scaling types: reactive, where resources adjust based on real-time changes, and proactive, which predicts future workloads for preemptive resource allocation.

As networks become more complex, more sophisticated techniques are needed. Recently, \ac{ml} strategies have been proposed for predictive scaling, using historical infrastructure metrics to forecast future demand and proactively make scaling decisions~\cite{duc2019machine, subramanya2021centralized, martin2022dimensioning}. Additionally, recent research has explored various \ac{rl}-based techniques to address the complex correlations between \ac{sla}, \ac{qos} parameters, scaling decision triggers, and learning metrics, which are often difficult to model accurately~\cite{gari2021reinforcement}.

For example, \citeauthor{rossi2019horizontal}~\cite{rossi2019horizontal} used \ac{rl} for horizontal and vertical scaling in docker swarm environment, showing that model-based approaches adeptly learn the optimal adaptation policy following user-defined deployment objectives. \citeauthor{he2021towards}~\cite{he2021towards} combined \ac{rl} with \acp{gnn} for chain-aware scaling in network services, outperforming traditional methods in cost and efficiency. The reasons behind the improvement are based on the prototype's ability to i) learn from past experiences, ii) better demand prediction thanks to the composite features, and iii) the incorporation of the global chain information into scaling decisions.

Similarly, \citeauthor{khaleq2021intelligent}~\cite{khaleq2021intelligent} used \ac{rl} to enhance \ac{k8s} auto-scaling, reducing response times by 20\%. The \ac{rl} algorithm identifies the right threshold values for pod scaling, which are then fed to the \ac{k8s} \ac{hpa}.

\citeauthor{soto2021towards}~\cite{soto2021towards, soto2023network} compared \ac{rl} to control-based methods, showing \ac{rl}’s flexibility but noting that it heavily depends on reward function design, where light variations of the \ac{rfn} result in different behaviors, while the control-theory based scaler is more deterministic in the scaling decisions. Lastly, \citeauthor{santos2023gym}~\cite{santos2023gym} developed the gym-hpa framework to train \ac{rl} agents in real cloud environments, achieving significant reductions in resource usage by at least 30\% and application latency by 25\% compared to traditional scaling methods.

\subsection{Benchmarks in Reinforcement Learning}
In \ac{ml}, a benchmark refers to a standardized set of tasks, datasets, and performance metrics used to evaluate and compare the performance of various algorithms. It ensures fair comparisons and tracks advancements in different fields. Benchmarks are widely used in \ac{ml} to measure algorithmic advancements, facilitate fair comparisons, and track the state-of-the-art in different subfields.

\ac{ml} benchmarks typically include tasks, curated datasets, performance metrics like accuracy or F1 score, evaluation protocols, and baseline models for reference. For \ac{sl} tasks, such as classification and regression,  benchmarks often include standardized datasets for training, validation, and testing~\cite{deng2009imagenet,tay2020long, warden2018speech}. Conversely, \ac{rl} employs standard environments, like OpenAI Gym~\cite{brockman2016openai}, for evaluation. \ac{rl} tasks vary, from discrete action spaces in Atari games to continuous control problems using physics engines like MuJoCo.

However, reproducibility in \ac{rl} is challenging~\cite{henderson2018deep, islam2017reproducibility}. Studies have shown that different implementations of the same RL algorithm can yield varied results, due to differences in hyperparameter settings and the lack of standardized reporting. Ensuring reliable comparisons requires multiple trials with different random seeds and proper significance testing.

In networking, standard implementation of \ac{rl} algorithms are still emerging. An initial step involves establishing an abstraction layer that enables the representation of a network environment as a Gym environment. This layer exposes simulation entities' states and control parameters for the agent's learning objectives. Tools like \textit{ns3-gym}~\cite{gawlowicz2019ns, yin2020ns3} integrate network simulators like ns-3~\cite{riley2010ns} with \ac{ml} frameworks such as PyTorch and TensorFlow for building \ac{rl} applications.

Improvements in communication methods have made \ac{rl}-based networking applications more efficient. Recent advancements include \textit{OpenRAN Gym}~\cite{bonati2023openran, bonati2021colosseum}, which combines several software frameworks for \ac{ran} data collection and control in 5G networks, demonstrating the viability of transferring trained models from simulations to real-world platforms.

\subsection{Differences with previous works}
In previous subsections, we i) reviewed state-of-the-art techniques for scaling with special emphasis in \ac{rl} approaches and ii) described common benchmarks in \ac{rl} and reviewed the current body of literature regarding the integration of networking and \ac{rl}. 

On the one hand, the works showing \ac{rl} techniques for scaling are opaque in reporting the work done at fine-tuning their \ac{rl} agents. They lack a detailed overview of the parameters used in their agents, only show the results of their best-performing agent, or do not specify how many runs of the same algorithm they performed, a practice highly discouraged by \ac{rl} practitioners since it hampers the reproducibility of such techniques and further development. An exception to this common practice in networking is our previous work~\cite{soto2023network}, where we showed the impact of the \ac{rfn} definition on the performance of the \ac{rl} agent and the variability of its performance even using the same \ac{rfn}. 

On the other hand, there are growing efforts in designing frameworks that allow a common and standard way of training and evaluating \ac{rl} algorithms for networking. Nevertheless, such efforts focus more on \ac{ran} implementations than on cloud-like environments. The ns-3 simulator puts more effort into simulating actual devices but lacks support for service-based network architectures. Several extensions are available but tailored to specific cases, such as \ac{fpga} implementations for 5G networks~\cite{miozzo2018sdr} and \ac{oran} solutions~\cite{garey2023ran}, where the service management and orchestration module, in charge of the xApps lifecycle operations (scaling being one of them), is out of the scope of their model. Moreover, the work proposed by \textit{OpenRAN Gym} is interesting since it created a platform to train and evaluate \ac{ml} and \ac{rl} models using large-scale emulators. However, since their interfaces do not follow the OpenAI Gym abstractions, it limits the benchmarking of \ac{rl} algorithms and compatibility with future \ac{rl} frameworks in other network domains.

With this work, we pretend to tackle the deficiencies mentioned above. Concerning the reproducibility issue, we thoroughly describe our training and testing procedures. Moreover, we follow the standard implementations of stable baselines, which already deliver the tuned hyper-parameters as intended in their original versions. Furthermore, we open-source our environment and abstraction layer for scaling in service-based network architectures, similar to the work done in~\cite{santos2023gym}. Though, being a simulator, \emph{DynamicSim} avoids the impracticality of owning or renting a \ac{k8s} cluster, accelerating the learning curve of new \ac{rl} in networking practitioners. We hope this work will foster research in the applicability of \ac{rl} in networking and further adoption in real-life environments. 
\section{Conclusion and future research directions} \label{sec:conclusion}

\ac{ml} algorithms are transforming 6G networks by optimizing areas such as resource management, security, and user experience. As these algorithms become integral to network operations, the need for efficient performance evaluation using relevant \acp{kpi} grows. This paper introduces a methodology for designing, training, and evaluating \ac{drl} algorithms, focusing on scaling resources in service-based networks. The scaling problem is modeled as an \ac{mdp}, where the goal is to autonomously determine the number of replicas in a changing environment. The study tested different \acp{rfn} to guide agent learning. Following common practices in the \ac{rl} community, we experimentally evaluated the behavior of the \ac{drl} agents. We determined, when possible, the best-performing agent in terms of the replica creation and latency control. It was observed that only \ac{rfn}1 showed a consistent statistical difference across the training epochs, with the \ac{dqn} algorithm outperforming \ac{ppo}. However, more data is needed to draw conclusions for other \acp{rfn}. 

The paper also highlights the challenge of fairly comparing scaling algorithms, proposing a methodology that integrates learning and performance metrics like replica creation and \ac{sla} compliance. This methodology helps in performing autonomous model selection, a crucial step towards completely autonomous network operation, where ideally, the \ac{nio} should select the best-performing algorithm for this task. In that sense, our methodology extends current \ac{mlops} frameworks by considering multiple models designed with different \acp{rfn} performing the same task.

Based on the literature analysis, we observe that the applicability of \ac{rl} techniques in networking leave ample room for further innovation. Thus, as a contribution to closing that gap, we provided an abstraction layer to integrate two platforms, \textit{OpenAI Gym}, the most used library to train and evaluate \ac{rl} algorithms, and \textit{DynamicSim}, a discrete-event simulator that enables the creation of edge-cloud network scenarios. With this abstraction layer, we created the playground on which scaling algorithms based on \ac{rl} can be freely trained, tested, and compared. 

Future work includes extending \textit{DynamicSim} to handle more complex scenarios, such as service function chaining and energy consumption optimization, and addressing the challenges of real-world implementations.



\printcredits

\section*{Funding sources}
The European Union partially funds this research under Grant Agreements No.~101017109 (DAEMON - Horizon 2020) and No.~101136314 (6G-TWIN - SNS). The views expressed are those of the authors and do not necessarily represent the views of the European Union or Smart Networks and Services Joint Undertaking. Additionally, the work is supported by the imec.icon project 5GECO (HBC.2021.0673), co-financed by imec and receiving financial backing from Flanders Innovation \& Entrepreneurship and Innoviris.

\bibliographystyle{cas-model2-names}
\bibliography{cas-refs.bib}

\section*{Biographies}
\bio{}
\textbf{Paola Soto} is a Ph.D. researcher at the University of Antwerp - imec. She received her B.Sc. in Electronics and her M.Sc. in Telecommunications Engineering from the University of Antioquia, Colombia, in 2014 and 2018, respectively. Her current research is focused on developing network management strategies using artificial intelligence and machine learning.
\endbio


\bio{}
\textbf{Miguel Camelo, Ph.D.} received a master’s degree in systems and computer engineering (University of Los Andes, Colombia, 2010) and a Ph.D. degree in computer engineering (University of Girona, Spain, 2014). He has authored several papers in international conferences/journals. He is a Senior Researcher at the University of Antwerp - imec, Belgium, where he leads the research on applied artificial intelligence (AI) in networking. His research interests are in the field of applied AI in communication networks.
\endbio

\bio{}
\textbf{Danny De Vleeschauwer, Ph.D.} received the M.Sc. degree in electrical engineering and the Ph.D. degree in applied sciences from Ghent University, Belgium, in 1985 and 1993, respectively. He currently is a principal research engineer in the Network Automation Department of the Network Systems and Security Research Lab of Nokia Bell Labs in Antwerp, Belgium. Before joining Nokia, he was a Researcher at Ghent University. His early work was on image processing and the application of queuing theory in packet-based networks. His current research interest includes the distributed control of applications over packet-based networks. 
\endbio

\bio{}
\textbf{Yorick De Bock, Ph.D.} obtained his Doctor Degree in Applied Engineering at the University of Antwerp on hard real-time virtualization for multi-core embedded systems. Yorick is a member of the IDLab research group, a joint research initiative between the University of Antwerp and Ghent University, and a core research group of imec. Currently, he is working as a software developer focusing on making prototypes for multiple IoT and AI projects.
\endbio

\bio{}
\textbf{Nina Slamnik-Kriještorac, Ph.D.} is a principal investigator at imec Research Center in Belgium and the University of Antwerp. In 2016, she obtained her Master's degree in telecommunications engineering at the Faculty of Electrical Engineering, University of Sarajevo, Bosnia \& Herzegovina. Nina obtained her Ph.D. in 2022, at the University of Antwerp. Nina has published numerous research papers in top-tier conferences and journals, and she is currently active in several European projects that are creating 5G enhancements for the transport \& logistics sector. Her current research focuses on developing zero-touch techniques for optimizing and automating the management and orchestration of EdgeApps within 6G ecosystems.
\endbio

\bio{}
\textbf{Chia-Yu Chang, Ph.D.} received his Ph.D. from Sorbonne Université, France, and is currently a senior research engineer at Nokia Bell Labs, Belgium. He has more than 12 years of experience in algorithm/protocol research on communication systems and network applications in academic and industrial laboratories, including EURECOM Research Institute, MediaTek, Huawei Swedish Research Center, and Nokia Bell Labs. His research interests include wireless communication, computer networking, low-latency low-loss scalable throughput (L4S), and AI/ML-supported network control.
\endbio

\bio{}
\textbf{Natalia Gaviria, Ph.D.} is an associate professor at the Electronics and Telecommunications Engineering Department at the University of Antioquia, Medellín, Colombia. In 1996 she received her BSc. Eng in Electronics Engineering from the University of Antioquia; in 1999, she received her MSc. degree in Electrical Engineering from the University of Los Andes, Colombia, and in 2006, she received her Ph.D. in Computers and Electrical Engineering from The University of Arizona, Tucson, USA. Her research interests include traffic theory, modeling of wireless networks and technical aspects application of Wireless Technology in telemedicine.
\endbio

\bio{}
\textbf{Erik Mannens, Ph.D.} is Director @ imec UAntwerp IDLab \& Professor \@ UAntwerp (Sustainable AI) and @ Ghent University (Semantic Intelligence). Since 2005 he has successfully managed +160 "interdisciplinary" projects (amounting 30M euro of Funding for his team) and teams of 50 to 125 researchers. He received his PhD degree in Computer Science Engineering (2011) at UGent and his Master’s degree in Computer Science (1995) at K.U. Leuven University \& his Master’s degree in Electro-Mechanical Engineering (1992) at KAHO Ghent. 
\endbio

\bio{}
\textbf{Juan F. Botero, Ph.D.} received the Ph.D. degree in telematics engineering from the Technical University of Catalonia, Spain, in 2013. He is an Electronics and Telecommunications Engineering Department associate professor at Universidad de Antioquia, Colombia (UdeA). In 2013, he joined the GITA Lab research group at UdeA. His main research interests include quality of service, Software Defined Networks, NFV, cybersecurity, network management, and resource allocation.
\endbio

\bio{}
\textbf{Steven Latré, Ph.D.} heads imec’s artificial intelligence research. He joined imec in 2013. Steven also headed the IDLab research group, which he helped grow to over 100 members. He received a Ph.D. in computer science engineering from Ghent University, Belgium, in 2011. He has authored over 100 papers in international journals/conferences. He is a recipient of the IEEE COMSOC Award for the Best Ph.D. in Network and Service Management (2012), the IEEE NOMS Young Professional Award (2014), the IEEE COMSOC Young Professional Award (2015), and the Laureate of the Belgian Academy (2019). His main expertise focuses on combining sensor technologies and chip design with AI to provide end-to-end solutions in various sectors.
\endbio

\end{document}